\documentclass[%
 reprint,
superscriptaddress,
 amsmath,amssymb,
 aps,
]{revtex4-2}
\usepackage{subcaption}
\usepackage{graphicx}% Include figure files
\usepackage{dcolumn}% Align table columns on decimal point
\usepackage{bm}% bold math
\begin{document}

\preprint{APS/123-QED}

\title{Degenerate squeezing in a dual-pumped integrated microresonator: parasitic processes and their suppression}% Force line breaks with \\
\author{H. Seifoory} 
\email{hossein.seifoory@utoronto.ca} 
\affiliation{Department of Physics, University of Toronto, 60 St. George Street, Toronto, Ontario M5S 1A7, Canada}
\affiliation{Xanadu, Toronto, ON, M5G 2C8, Canada}
\author{Z. Vernon}
\author{D. H. Mahler}
\author{M. Menotti}
\author{Y. Zhang}
\affiliation{Xanadu, Toronto, ON, M5G 2C8, Canada}
\author{J. E. Sipe}
\affiliation{Department of Physics, University of Toronto, 60 St. George Street, Toronto, Ontario M5S 1A7, Canada}

%\address{$^{1}$Department of Physics, University of Toronto, 60 St. George Street, Toronto, Ontario M5S 1A7, Canada}
%\address{$^{2}$Xanadu, Toronto, ON, M5G 2C8, Canada}

\date{\today}% It is always \today, today,

\begin{abstract}
Using a general Hamiltonian treatment, we theoretically study the generation of degenerate quadrature squeezing in a dual-pumped integrated microring resonator coupled to a waveguide. Considering a dual-pump four-wave mixing configuration in an integrated $\text{Si}_3\text{N}_4$ platform, and following the coupled-mode theory approach, we investigate the effects of parasitic quantum nonlinear optical processes on the generation of squeezed light. Considering five resonance modes in this approach allows us to include the most important four-wave mixing processes involved in such a configuration. We theoretically explore the effects of the pump detunings on different nonlinear processes and show that the effects of some of the parasitic processes are effectively neutralized by symmetrically detuning the two pumps. This yields a significant enhancement in the output squeezing quality without physically changing the structure, but suffers from the trade-off of requiring substantially higher pump power for a fixed target level of squeezing. 

\end{abstract}

\maketitle

%\tableofcontents

\section{\label{sec:introd}Introduction}
Squeezed states of light, in which the quantum fluctuations in one quadrature component are suppressed to a level below the quantum noise limit, have a wide range of applications in quantum metrology~\cite{PhysRevD.23.1693,Taylor2013,Otterstrom:14,Oelker:14}, quantum imaging~\cite{PhysRevLett.85.3789, PhysRevLett.88.203601,PhysRevLett.105.053601}, and communication~\cite{PhysRevLett.80.869,Furusawa706}. In addition, continuous variable~(CV) entanglement can be efficiently produced using squeezed light and linear optics~\cite{PhysRevLett.82.1784,lpor.200910010,RevModPhys.77.513}. Of the methods used to generate squeezed states~\cite{PhysRevLett.47.709,PhysRevLett.59.198,PhysRevLett.81.3635,PhysRevLett.109.013601}, the most common involve the use of either spontaneous four-wave mixing~(SFWM)~\cite{PhysRevLett.55.2409} or spontaneous parametric down conversion~(SPDC)~\cite{PhysRevLett.57.2520}. Implementations in bulk optics can suffer from scalability and complexity issues, but with the development of integrated photonics technology it is possible to overcome these limitations by integrating  squeezed-light sources on one monolithic platform~\cite{PhysRevApplied.12.064024, Hoff:15,PhysRevApplied.3.044005, Vaidyaeaba9186}. In particular, squeezed light generation via single-pass waveguides has been reported~\cite{Mondain:19}, and nonlinear optical resonators such as micro-ring resonators, whispering gallery mode resonators, and coupled-resonator optical waveguides can be employed to enhance the squeezing of the generated light~\cite{Clemmen:09,doi:10.1002/lpor.201100017,PhysRevA.91.053802,DelHaye2007,Lin:17,Ferrera2008,Melloni2003,PhysRevA.100.033839,PhysRevA.95.033815,PhysRevA.97.023840}.\par  
Many CV protocols, such as CV quantum sampling\cite{PhysRevApplied.12.064024}, require squeezing in a single mode. One effective strategy for achieving this is based on dual-pump SFWM in microring resonators~\cite{PhysRevApplied.12.064024,Zhang2021}. In general, the SFWM process involves the conversion of two pump photons to signal and idler photons~\cite{boyd2019nonlinear}, where from energy conservation we have $\hbar \omega_{P_1} + \hbar \omega_{P_2} \ = \hbar \omega_{S} + \hbar \omega_{I}$, with $\omega_{P_1}$ and $\omega_{P_2}$ the two input frequencies, and $\omega_{S}$ and $\omega_{I}$ the generated signal and idler frequencies. For squeezing in a single mode, a prerequisite is to be in the degenerate squeezing regime, where the signal and idler frequencies are within the same resonance. However, other concurrent processes can lead to parasitic effects polluting the squeezed light~\cite{PhysRevLett.118.073603,PhysRevLett.124.193601,Zhang2021}. The study of such parasitic processes is essential to the identification of proper suppression strategies.\par    
In this work, we consider a ring resonator, made of a third-order nonlinear optical material such as silicon nitride~($\text{Si}_3 \text{N}_4$), dual-pumped through a side-coupled waveguide. The optical properties of such a structure, schematically shown in Fig.~\ref{fig:maximums_not_equal_sigma_appendix}a, can be studied theoretically using either the Lugiato-Lefever equations (LLE) or coupled mode theory (CMT). The former, which is a spatio-temporal method, requires the solution of only a single partial differential equation, and hence is computationally cost effective. However, it does not allow for an individual investigation of the consequences of each frequency mixing process. In contrast, for low enough input pump powers one can employ CMT, which is a spectro-temporal method, limited to only five resonances; we label them as $\{m,p_{1},s,p_{2},n\}$. As shown in Fig.~\ref{fig:processes}, $p_{1}$ and $p_{2}$ are the two resonances used for pumping, $s$ is the desired resonance for the squeezed light, and $m$ and $n$ are the adjacent resonances of $p_{1}$ and $p_{2}$, respectively. In this paper we use the CMT method to identify the role that the different frequency mixing processes play in determining the squeezing and anti-squeezing of light in resonance $s$, restricting ourselves to pump powers where an LLE analysis confirms that the restriction to five resonances is a reasonable approximation. This extends previous studies, where the calculations were limited to only three resonances, or only considered a single pump ~\cite{PhysRevLett.122.153906,PhysRevA.91.053802}. The dual-pump configuration allows us to study degenerate squeezing, and the inclusion of the two adjacent resonances enables an investigation of the effects of parasitic processes on the squeezing.\par
The paper is organized as follows. In Secs.~II-V we introduce the Hamiltonian terms associated with the ring fields, the channel fields, the coupling of the ring and channel, and the nonlinearity.  With advances in both the fabrication of integrated photonic structures and the measurement of their nonlinear quantum optical properties, a careful identification of the parameters that arise in the dynamical equations, which we present here, is in order. The correlation functions of interest are presented in Sec. VI, the dynamical equations that must be solved in Sec. VII, and the operator dynamics in Sec. VIII.  The results for a sample structure are given in Sec. IX, and we conclude in Sec. X.  Some of the calculation details are relegated to three appendices.    

\section{Ring fields}

\begin{figure}[!ht]
\begin{center}
\includegraphics[width=\linewidth]{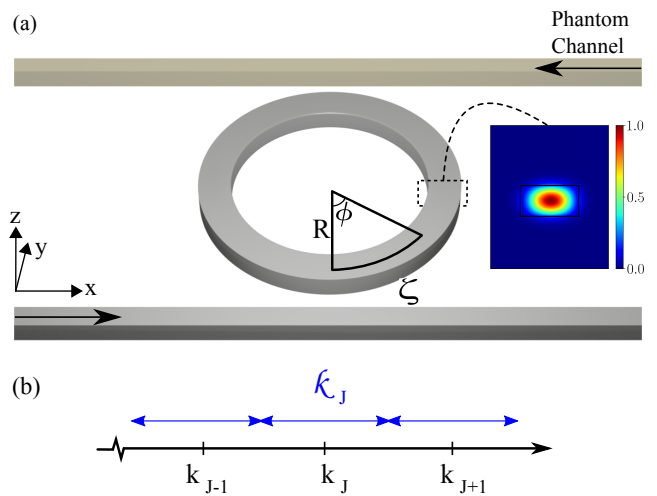}\
\caption{(a) Schematic of microring resonator, with the nominal radius $R$, side-coupled to a physical channel and an effective phantom channel, which is introduced to account for scattering losses. The $x$ coordinate for the phantom channel increases in the opposite direction to that of the real channel. The inset shows the electric field intensity of the fundamental mode at $1550\,\text{nm}$ inside the ring. The parameters for the ring and channel are given in section X. (b) The range $\mathcal{K}_{J}$ of $k$, given by Eq.~(\ref{eq:RJ}), associated with the $J^{th}$ ring resonance  }\label{fig:maximums_not_equal_sigma_appendix}
\end{center}
\end{figure}

We begin by considering an isolated ring resonator as shown in Fig.~\ref{fig:maximums_not_equal_sigma_appendix}a, but imagined far from any channel.  A description that respects the symmetry uses the cylindrical
coordinates $z$, the angle $\phi$, and the radial coordinate $\rho=\sqrt{x^{2}+y^{2}}$; see Fig.~\ref{fig:maximums_not_equal_sigma_appendix}. 
It is convenient to introduce a nominal radius $R$ and use $\zeta=R\phi$
as a coordinate in place of $\phi$. Denoting $\boldsymbol{r}_{\perp}=(\rho,z)$,
a volume element is $d\boldsymbol{r}=d\boldsymbol{r}_{\perp}d\zeta$,
where $d\boldsymbol{r}_{\perp}=R^{-1}\rho d\rho dz$, and $\zeta$
varies from $0$ to $\mathcal{L}\equiv2\pi R$. We assume that the
linear response of the structure can be described by a relative dielectric
constant $\varepsilon(\boldsymbol{r},\omega$), and from the symmetry
of the system the relative dielectric constant depends only on $\boldsymbol{r}_{\perp}$,
$\varepsilon(\boldsymbol{r},\omega)=\varepsilon(\boldsymbol{r_{\perp}},\omega)$.
We label the modes of the ring by $\kappa_{J}$ , the wave number
associated with their propagation along the ring; we have $\kappa_{J}=2\pi n_{J}/\mathcal{L}$,
where $n_{j}$ is an integer, and here we consider $n_{J}>0$. We
consider only one transverse field structure relevant for each $\kappa_{J}$,
and identify its frequency by $\omega_{J}$; this could be easily
generalized. The Hamiltonian for the modes in the ring is then of
the standard form, 
\begin{align}
 & H_{ring}=\sum_{J}\hbar\omega_{J}b_{J}^{\dagger}b_{J},
\end{align}
neglecting the zero-point energy, as we do throughout; here 
\begin{align}
 & \left[b_{J},b_{J'}^{\dagger}\right]=\delta_{JJ'}
\end{align}
as usual. It is convenient to specify the electromagnetic field amplitude
of each mode in terms of the displacement field and magnetic field~\cite{PhysRevA.91.053802}, in particular writing 
\begin{align}
 & \boldsymbol{D}(\boldsymbol{r})=\sum_{J}\sqrt{\frac{\hbar\omega_{J}}{2}}b_{J}\boldsymbol{D}_{J}(\boldsymbol{r})+H.c.,\label{eq:Ddef}
\end{align}
where 
\begin{align}
 & \boldsymbol{D}_{J}(\boldsymbol{r})=\frac{\mathsf{d}_{J}(\boldsymbol{r}_{\perp};\zeta)e^{i\kappa_{J}\zeta}}{\sqrt{\mathcal{L}}}.
\end{align}
The dependence of $\mathsf{d}_{J}(\boldsymbol{r}_{\perp};\zeta)$ on $\zeta$  arises due to its components in the $xy$ plane. As $\zeta$ varies they will lead to a change in direction of $\mathsf{d}_{J}(\boldsymbol{r}_{\perp};\zeta)$, despite the fact that $\mathsf{d}_{J}^{*}(\boldsymbol{r_{\perp}};\zeta)\cdot\mathsf{d}_{J}(\boldsymbol{r}_{\perp};\zeta)$ will be independent of $\zeta$; we can then take $\mathsf{d}_{J}^{*}(\boldsymbol{r_{\perp}};\zeta)\cdot\mathsf{d}_{J}(\boldsymbol{r}_{\perp};\zeta)=\mathsf{d}_{J}^{*}(\boldsymbol{r_{\perp}};0)\cdot\mathsf{d}_{J}(\boldsymbol{r}_{\perp};0)$.  We can include dispersion effects in the normalization of each
mode~\cite{PhysRevA.73.063808, Sipe_2009} by taking
\begin{align}
 & \int\int\frac{\mathsf{d}_{J}^{*}(\boldsymbol{r_{\perp}};0)\cdot\mathsf{d}_{J}(\boldsymbol{r}_{\perp};0)}{\epsilon_{0}\varepsilon(\boldsymbol{r}_{\perp},\omega_{J})}\frac{v_{p}(\boldsymbol{r}_{\perp},\omega_{J})}{v_{g}(\boldsymbol{r}_{\perp},\omega_{J})}d\boldsymbol{r}_{\perp}=1.\label{eq:normalization}
\end{align}
where $v_{p}(\boldsymbol{r}_{\perp},\omega_{J}$) and $v_{g}(\boldsymbol{r}_{\perp},\omega_{J}$)
are respectively the local phase and group velocities at the frequency
$\omega_{J}$ of the mode; in general 
\begin{align}
 & v_{p}(\boldsymbol{r_{\perp}},\omega)=c/n(\boldsymbol{r_{\perp}},\omega),\\
 & v_{g}(\boldsymbol{r}_{\perp},\omega)=v_{p}(\boldsymbol{r}_{\perp},\omega)\left[1+\frac{\omega}{n(\boldsymbol{r}_{\perp},\omega)}\frac{\partial n(\boldsymbol{r}_{\perp},\omega)}{\partial\omega}\right]^{-1},
\end{align}
where $n(\boldsymbol{r}_{\perp},\omega)=\sqrt{\varepsilon(\boldsymbol{r}_{\perp},\omega)}$
is the local index of refraction.
\section{Channel fields}

We next consider an isolated channel, as shown in the lower part of Fig.~\ref{fig:maximums_not_equal_sigma_appendix}a, but imagined far from any ring. Here we
take the channel to range along the $x$ coordinate, and now use $\boldsymbol{r}_{\perp}=(y,z)$. 
Considering only one transverse field structure associated with each
wave number $k$, in analogy with the treatment of the ring resonator,
the displacement field operator can be written~\cite{PhysRevA.91.053802}
\begin{align}
 & \boldsymbol{D}(\boldsymbol{r})=\int\sqrt{\frac{\hbar\omega_{k}}{4\pi}}a(k)\boldsymbol{d}_{k}(\boldsymbol{r}_{\perp})e^{ikx}dk+H.c.,\label{eq:DCexpand}
\end{align}
where $\omega_{k}$ is the frequency of the field characterized by
$k$, the integral is over the range of $k$ for which such modes
exist, and 
\begin{align}
 & \left[a(k),a^{\dagger}(k')\right]=\delta(k-k').
\end{align}
The $\boldsymbol{d}_{k}(\boldsymbol{r}_{\perp})$ satisfy the
normalization condition (\ref{eq:normalization}), with $\mathsf{d}_{J}(\boldsymbol{r}_{\perp};0)$
replaced by $\boldsymbol{d}_{k}(\boldsymbol{r}_{\perp}$) with
$\boldsymbol{r}_{\perp}=(y,z)$. The Hamiltonian of the channel~\cite{PhysRevA.91.053802} is given by
\begin{align}
 & H_{channel}=\int\hbar\omega_{k}a^{\dagger}(k)a(k)dk.\label{eq:HCexpand}
\end{align}
It is useful to break up the range of $k$ in (\ref{eq:DCexpand},\ref{eq:HCexpand})
into portions associated with each ring resonance. Letting $k_{J}$
be the value of $k$ for which $\omega_{k_{J}}=\omega_{J}$, the frequency
of the $J^{th}$ ring resonance, we set the range $\mathcal{K}_{J}$ of $k$
associated with ring resonance $J$ to be the values of $k$ for which
\begin{align}
 & k_{J}-\frac{(k_{J}-k_{J-1})}{2}<k\leq k_{J}+\frac{(k_{J+1}-k_{J})}{2},\label{eq:RJ}
\end{align}
(see Fig.~\ref{fig:maximums_not_equal_sigma_appendix}b); then we write 
\begin{align}
 & \boldsymbol{D}(\boldsymbol{r})=\sum_{J}\int_{\mathcal{K}_{J}}\sqrt{\frac{\hbar\omega_{k}}{4\pi}}a(k)\boldsymbol{d}_{k}(\boldsymbol{r}_{\perp})e^{ikx}dk+H.c.\label{eq:Dapprox}\\
 & \approx\sum_{J}\sqrt{\frac{\hbar\omega_{J}}{2}}\boldsymbol{d}_{J}(\boldsymbol{r}_{\perp})e^{ik_{J}x}\psi_{J}(x)+H.c.,\nonumber 
\end{align}
where we have put $\boldsymbol{d}_{J}(\boldsymbol{r}_{\perp})\equiv\boldsymbol{d}_{k_{J}}(\boldsymbol{r}_{\perp})$
and 
\begin{align}
 & \psi_{J}(x)=\int_{\mathcal{K}_{J}}a(k)e^{i(k-k_{J})x}\frac{dk}{\sqrt{2\pi}}.\label{eq:psiJdef}
\end{align}
In the second line of (\ref{eq:Dapprox}) we have neglected the variation
of $\omega_{k}$ and $d_{k}(\boldsymbol{r}_{\perp})$ as $k$ varies
over the range (\ref{eq:RJ}), and approximated them as $\omega_{J}$
and $\boldsymbol{d}_{J}(\boldsymbol{r}_{\perp})$ respectively. Clearly
$\psi_{J}(x)$ commutes with $\psi_{J'}(x')$ and $\psi_{J'}^{\dagger}(x')$
for $J'\neq J$; if the integral in (\ref{eq:psiJdef}) ranged from
$-\infty$ to $\infty$ we would have 
\begin{align}
 & \left[\psi_{J}(x),\psi_{J}^{\dagger}(x')\right]=\delta(x-x').
\end{align}
This is not exact, of course, but for fields $\psi_{J}(x)$ with relevant
components $k$ centered near $k_{J}$ and far from the boundaries
of $\mathcal{K}_{J}$ specified by (\ref{eq:RJ}) it will serve as a good approximation,
and we adopt it. Breaking the integral in the Hamiltonian (\ref{eq:HCexpand})
into different resonance ranges $\mathcal{K}_{J}$, and neglecting group-velocity dispersion within
each range by writing $\omega_{k}=\omega_{J}+v_{J}(k-k_{J})$ in range
$\mathcal{K}_{J}$, where $v_{J}$ is the group velocity of the channel field
at $k_{J}$, we have 
\begin{align}
 & H_{channel}=\sum_{J}\int_{\mathcal{K}_{J}}\hbar(\omega_{J}+v_{J}(k-k_{J}))a^{\dagger}(k)a(k)dk\label{eq:Hchanneluse}\\
 & =\sum_{J}\hbar\omega_{J}\int\psi_{J}^{\dagger}(x)\psi_{J}(x)dx\nonumber \\
 & -\frac{1}{2}i\hbar\sum_{J}v_{J}\int\left(\psi_{J}^{\dagger}(x)\frac{\partial\psi_{J}(x)}{\partial x}-\frac{\partial\psi_{J}^{\dagger}(x)}{\partial x}\psi_{J}(x)\right)dx.\nonumber 
\end{align}
Finally, the operator for the total power flow in the waveguide is given by
\begin{align}
 & P(x)=\int\boldsymbol{S}(\boldsymbol{r})\cdot\boldsymbol{\hat{x}}\;d\boldsymbol{r}_{\perp},
\end{align}
where $\boldsymbol{S}(\boldsymbol{r})$ is the Poynting vector operator.
Omitting terms that will be rapidly varying when we move to a Heisenberg
picture, we find~\cite{J_E_Sipe_2016}
\begin{align}
 & P(x)\rightarrow\sum_{J}\hbar\omega_{J}v_{J}\psi_{J}^{\dagger}(x)\psi_{J}(x).\label{eq:power}
\end{align}

\section{Ring-channel coupling}

We now adopt an effective point-coupling model between the channel and
the ring (see Fig.~\ref{fig:maximums_not_equal_sigma_appendix}a). If we take $\zeta=0$ to correspond to the
location on the ring closest to the channel, then the field in the
ring at that point will include contributions from all the $b_{J}$
with prefactors that will vary little as we move from one $J$ to
the next. Taking $x=0$ to be the point in the channel closest to
the ring, the Hamiltonian for an effective point-coupling model~\cite{PhysRevA.91.053802}
is then 
\begin{align}
 & H_{coupling}=\sum_{J}\left(\hbar\gamma_{J}^{*}b_{J}^{\dagger}\psi(0)+H.c.\right),
\end{align}
where we expect the coupling constants $\gamma_{J}$ to vary little
as we move from one $J$ to the next.\par  Most of the nonclassical attributes of light, such as squeezing and entanglement, are fragile with respect to loss~\cite{Jasperse:11,josab-34-8-1587,Ulanov2015}; this is an important issue when resonant structures are being studied. To take into account scattering
losses in the ring we adopt a beam-splitter approach~\cite{PhysRevA.91.053802} by introducing
a ``phantom'' channel (Fig.~\ref{fig:maximums_not_equal_sigma_appendix}a) with a coupling to the ring given
by 
\begin{align}
 & H_{coupling}^{ph}=\sum_{J}\left(\hbar\gamma_{Jph}^{*}b_{J}^{\dagger}\psi_{ph}(0)+H.c.\right),
\end{align}
where $\psi_{Jph}(x)$ is a field operator for the phantom channel, 
\begin{align}
 & \psi_{Jph}(x)=\int_{\mathcal{K}_{J}^{ph}}a_{ph}(k)e^{i(k-k_{Jph})x}\frac{dk}{\sqrt{2\pi}}
\end{align}
(cf. (\ref{eq:psiJdef})), with a phantom channel Hamiltonian 
\begin{align}
 H_{channel}^{ph}=& \sum_{J}\int_{\mathcal{K}_{J}^{ph}}\hbar(\omega_{J}+v_{Jph}(k-k_{Jph}))\nonumber\\
 &\times a^{\dagger}_{ph}(k)a_{ph}(k)dk,
\end{align}
(cf. (\ref{eq:Hchanneluse})), where $\gamma_{Jph},$ $a_{ph}(k)$,
$k_{Jph}$, and $v_{Jph}$ are the phantom channel analogs of the
quantities $\gamma_{J}$, $a(k)$, $k_{J}$ and $v_{J}$ characterizing
the actual channel and its coupling to the ring. We
take the initial state of the phantom channel to be vacuum, and note that we take the $x$ coordinate of the phantom channel to run in the opposite direction of that of the actual channel.

\section{Nonlinearity}

\begin{figure}
\includegraphics[width=\linewidth]{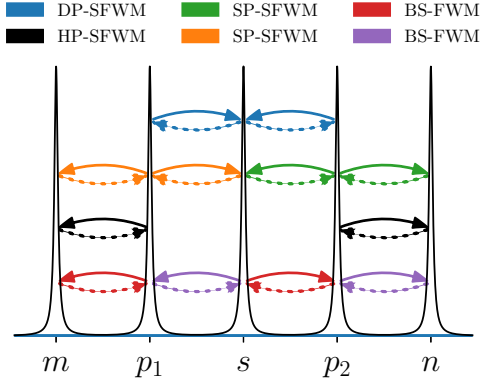}
\caption{(Color online) Four-wave mixing processes that occur in a ring resonator when two resonances $p_1$ and $p_2$ are pumped.  DP-SFWM leads to squeezing of the $s$ mode while the unwanted processes of SP-SFWM and BS-FWM generate noise and impurify the output. The corresponding conjugate of each process is shown with the dotted arrows with the same color.}
\label{fig:processes}
\end{figure}

For the $\chi_{(3)}^{ijkl}$ nonlinearity considered here, the nonlinear
Hamiltonian is given~\cite{PhysRevA.73.063808, PhysRevE.69.016604, PhysRevE.70.066621} by 
\begin{align}
 & H_{NL}=-\frac{1}{4\epsilon_{0}}\int d\boldsymbol{r}\Gamma_{(3)}^{ijkl}(\boldsymbol{r})D^{i}(\boldsymbol{r})D^{j}(\boldsymbol{r})D^{k}(\boldsymbol{r})D^{l}(\boldsymbol{r}),\label{eq:HNLdef}
\end{align}
where neglecting dispersion in this term we have 
\begin{align}
 & \Gamma_{(3)}^{ijkl}(\boldsymbol{r})=\frac{\chi_{(3)}^{ijkl}(\boldsymbol{r})}{\epsilon_{0}^{2}n^{8}(\boldsymbol{r})},\label{eq:gamma_def}
\end{align}
with $n(\boldsymbol{r})$ is the local index of refraction; we take
the integral in (\ref{eq:HNLdef}) to range over the volume of the
ring, where the nonlinear effect will be important. In this paper
we focus on five resonances of interest, taking the index $J$ in equations
such as (\ref{eq:Ddef}) to range over the labels $\left\{ m,p_{1},s,p_{2},n\right\} $
(see Fig.~\ref{fig:processes}). Here $p_{1}$ and $p_{2}$ label the two strong
pumps, leading to the generation of photons within resonances $m$, $s$, and $n$
by nonlinear processes we detail below; since the frequencies of these
five resonances are very close, the neglect of dispersion in (\ref{eq:gamma_def})
should be a good approximation. Using the displacement field (\ref{eq:Ddef})
in (\ref{eq:HNLdef}), and neglecting corrections due to normal ordering,
we find 
\begin{align}
 & H_{NL}=-\frac{\hbar}{2}\sum_{J,K,L,M}\Lambda_{JKLM}b_{J}^{\dagger}b_{K}^{\dagger}b_{L}b_{M},
\end{align}
where we have kept only the terms involving two creation and annihilation
operators, since the others will be rapidly varying in an interaction
picture and can be neglected, and
\begin{align}
\label{Lambda_JKLM_1}
 & \Lambda_{JKLM}=\frac{3\hbar}{4\epsilon_{0}}\sqrt{\omega_{J}\omega_{K}\omega_{L}\omega_{M}}\nonumber\\
 &\times \int d\boldsymbol{r}\Gamma_{(3)}^{ijkl}(\boldsymbol{r})\left(D_{J}^{i}(\boldsymbol{r})D_{K}^{j}(\boldsymbol{r})\right)^{*}D_{L}^{k}(\boldsymbol{r})D_{M}^{l}(\boldsymbol{r}).
\end{align}
where we have used (\ref{eq:Ddef}), and $\Delta\kappa=\kappa_L+\kappa_M-\kappa_J-\kappa_K=0$ for the energy conserving terms.  Again, since the frequencies of the five resonances of interest are close to each other, and thus to a reference frequency $\omega$, for the energy conserving terms that arise all the coefficients of $\Lambda_{JKLM}$ are essentially the same, and we can set $\Lambda_{JKLM}\rightarrow\Lambda$. In Appendix A we show that 
\begin{align}
 & \Lambda=\frac{\hbar\omega v^2\gamma}{\mathcal{L}},
\end{align}
where $v$ is the group velocity of light propagating in the ring, and $\gamma$ is the waveguide nonlinear parameter.  This corrects a factor of two error that appeared in a recent study~\cite{PhysRevApplied.12.064024}, and is in agreement with earlier work~\cite{zachary_thesis}. 

The relevant processes described
by $H_{NL}$, and their associated Hamiltonians, are then self-phase modulation (SPM), 
\begin{align}
H_{SPM}=-\hbar\frac{\Lambda}{2}\left(b_{p_{1}}^{\dagger}b_{p_{1}}^{\dagger}b_{p_{1}}b_{p_{1}}+b_{p_{2}}^{\dagger}b_{p_{2}}^{\dagger}b_{p_{2}}b_{p_{2}}\right),
\end{align}
and cross-phase modulation (XPM),
\begin{align}
H_{XPM} & =-2\hbar\Lambda\left(b_{s}^{\dagger}b_{p_{1}}^{\dagger}b_{s}b_{p_{1}}+b_{s}^{\dagger}b_{p_{2}}^{\dagger}b_{s}b_{p_{2}}+b_{p_{1}}^{\dagger}b_{p_{2}}^{\dagger}b_{p_{1}}b_{p_{2}}\right.\nonumber\\& \left.+b_{m}^{\dagger}b_{p_{1}}^{\dagger}b_{m}b_{p_{1}}+b_{m}^{\dagger}b_{p_{2}}^{\dagger}b_{m}b_{p_{2}}+b_{n}^{\dagger}b_{p_{1}}^{\dagger}b_{n}b_{p_{1}}\right.\nonumber\\
& +\left.b_{n}^{\dagger}b_{p_{2}}^{\dagger}b_{n}b_{p_{2}}\right),
\end{align}
which lead to frequency shifts of the ring resonances, together with dual-pump spontaneous four-wave mixing (DP-SFWM), 
\begin{align}
H_{DP-SFWM}=-\hbar \Lambda b_{s}^{\dagger}b_{s}^{\dagger}b_{p1}b_{p2}+H.c.,
\end{align}
single-pump SFWM (SP-SFWM),
\begin{align}
H_{SP-SFWM}=-\hbar\Lambda\left(b_{m}^{\dagger}b_{s}^{\dagger}b_{p_{1}}b_{p_{1}}+b_{n}^{\dagger}b_{s}^{\dagger}b_{p_{2}}b_{p_{2}}\right)+H.c.,
\end{align}
Bragg-scattering FWM (BS-FWM),
\begin{align}
H_{BS-FWM}=-2\hbar\Lambda\left(b_{p_{2}}^{\dagger}b_{m}^{\dagger}b_{p_{1}}b_{s}+b_{p_{1}}^{\dagger}b_{n}^{\dagger}b_{p_{2}}b_{s}\right)+H.c.,
\end{align}
and hyper-parametric SFWM,
\begin{align}
H_{HP-SFWM}=-2\hbar \Lambda b_{m}^{\dagger}b_{n}^{\dagger}b_{p1}b_{p2}+H.c.,
\end{align}
which are processes that lead to transitions between the different resonances. These processes, and the associated resonances involved, are illustrated in Fig.~\ref{fig:processes}. For each process the solid lines indicate the transitions that will be more important for the excitation scenario considered; the dotted lines, corresponding to the Hamiltonian terms denoted by $+H.c.$ in the expressions above, indicate the reverse transitions. For example, for DP-SFWM, which is the desired process for producing degenerate squeezed vacuum states in resonance $s$, the important transitions are those where pump photons from $p_1$ and $p_2$ are destroyed and two photons in resonance $s$ are created; the reverse transition, where two photons from resonance $s$ are destroyed and pump photons at $p_1$ and $p_2$ are created, will be less important since there will always be many more photons in $p_1$ and in $p_2$ than in $s$. 

Finally, then, we can then write 
\begin{align}
&H_{NL}\left(\left\{ b_{K}\right\} ,\left\{ b_{L}^{\dagger}\right\} \right)=  H_{DP-SFWM}+H_{SPM}+H_{XPM}\nonumber\\
&+H_{SP-SFWM}+H_{BS-FWM}+H_{HP-SFWM}.
\end{align}

 \par

\section{Correlation functions}

We now write our full Hamiltonian as 
\begin{align}
 & H=H_{0}+V(\left\{ \mathcal{A}_{i}\right\} ),
\end{align}
where 
\begin{align}
 & H_{0}=H_{ring}+H_{channel}+H_{channel}^{ph},\label{eq:Hamiltonian_start}\\
 & V(\left\{ \mathcal{A}_{i}\right\} )=H_{coupling}+H_{coupling}^{ph}+H_{NL}\left(\left\{ b_{J}\right\} ,\left\{ b_{L}^{\dagger}\right\} \right),\nonumber 
\end{align}
and the set $\left\{ \mathcal{A}_{i}\right\} $ consists of all the
Schrödinger operators appearing in $H_{coupling}$, $H_{coupling}^{ph},$
and $H_{NL}\left(\left\{ b_{J}\right\} ,\left\{ b_{L}^{\dagger}\right\} \right)$.
The evolution operators $\mathcal{U}_{0}(t,t')$ and $\mathcal{U}(t,t')$
for $H_{0}$ and $H(t)$, respectively, satisfy 
\begin{align}
 & i\hbar\frac{\partial}{\partial t}\mathcal{U}_{0}(t,t')=H_{0}\mathcal{U}_{0}(t,t'),\\
 & i\hbar\frac{\partial}{\partial t}\mathcal{U}(t,t')=H(t)\mathcal{U}(t,t'),
\end{align}
with $\mathcal{U}(t,t)=\mathcal{U}_{0}(t,t)=\mathbb{I}$ for all $t$.

For the Hamiltonians $H$ and the initial kets we consider, there
will be times $t_{min}$ and $t_{max}$ such that for all times $t<t_{min}$
and all times $t>t_{max}$ the term $V(\left\{ \mathcal{A}_{i}\right\} )$
will have no effect on the evolution of the Schrödinger ket $\left|\Psi(t)\right\rangle $.
We choose $t_{0}<t_{min}$ and $t_{1}>t_{max}$, and define 
\begin{align}
 & \left|\Psi_{in}\right\rangle =\mathcal{U}_{0}(0,t_{0})\left|\Psi(t_{0})\right\rangle ,\\
 & \left|\Psi_{out}\right\rangle =\mathcal{U}_{0}(0,t_{1})\left|\Psi(t_{1})\right\rangle .
\end{align}
Then, since $\left|\Psi(t_{1})\right\rangle =\mathcal{U}(t_{1},t_{0})\left|\Psi(t_{0})\right\rangle $
we have 
\begin{align}
 & \left|\Psi_{out}\right\rangle =U(t_{1},t_{0})\left|\Psi_{in}\right\rangle \label{eq:intoout}\\
 & =U(\infty,-\infty)\left|\Psi_{in}\right\rangle ,\nonumber 
\end{align}
where 
\begin{align}
 & U(t,t')\equiv\mathcal{U}_{0}(0,t)\mathcal{U}(t,t')\mathcal{U}_{0}(t',0),\label{eq:Udef}
\end{align}
and the extension to $\pm\infty$ in the second line of (\ref{eq:intoout})
follows because in the added times $V(\left\{ \mathcal{A}_{i}\right\} )$
has no effect. This is the approach of scattering theory. For our
applications in this paper we will be primarily interested in correlation
functions, and so a Heisenberg strategy, which is however connected
to this, will be more useful.

Defining Heisenberg operators referenced to $t_{0}$ in the usual
way, 
\begin{align}
 & \mathcal{A}_{i}^{H}(t)=\mathcal{U}^{\dagger}(t,t_{0})\mathcal{A}_{i}\mathcal{U}(t,t_{0}),
\end{align}
we use (\ref{eq:Udef}) to write $\mathcal{U}(t,t_{0})$ in terms
of $U(t,t_{0})$ and $\mathcal{U}_{0}$, and we find 
\begin{align}
 & \mathcal{A}_{k}^{H}(t)=\mathcal{U}_{0}^{\dagger}(0,t_{0})\left(\sum_{l}g_{kl}(t)\bar{\mathcal{A}}_{l}(t)\right)\mathcal{U}_{0}(0,t_{0}).\label{eq:Heisenberg}
\end{align}
Here 
\begin{align}
 & \bar{\mathcal{A}}_{l}(t)=U^{\dagger}(t,-\infty)\mathcal{A}_{l}U(t,-\infty),\label{eq:Abardef}
\end{align}
where we have used the fact that $U(t,t_{0})=U(t,-\infty)$, while
the $g_{kl}(t)$ are the functions that characterize an evolution
according to only $H_{0}$, 
\begin{align}
 & \mathcal{U}_{0}^{\dagger}(t,0)\mathcal{A}_{i}\mathcal{U}_{0}(t,0)=\sum_{j}g_{ij}(t)\mathcal{A}_{j}.\label{eq:gdef}
\end{align}
and we have taken the set $\left\{ \mathcal{A}_{i}\right\} $ to be
expanded, if necessary, so that these equations can be written for
the $\mathcal{A}_{i}$ originally appearing in $V\left(\left\{ \mathcal{A}_{i}\right\} ;t\right)$;
we will see examples of (\ref{eq:gdef}), and its extensions, below.
The $\bar{\mathcal{A}}_{l}(t)$ are easily seen to satisfy the equations
\begin{align}
 & i\hbar\frac{d}{dt}\bar{\mathcal{A}}_{l}(t)=\left[\bar{\mathcal{A}}_{l}(t),V\left(\left\{ \sum_{j}g_{ij}(t)\bar{A}_{j}(t)\right\} \right)\right],\label{eq:dynamical_equations}
\end{align}
with the initial conditions $\bar{\mathcal{A}}_{l}(-\infty)=\mathcal{A}_{l}$.
Using (\ref{eq:Heisenberg}) we then have 
\begin{align}
 \left\langle \Psi(t)|\mathcal{A}_{k}|\Psi(t)\right\rangle =& \left\langle \Psi(t_{0})|\mathcal{A}_{k}^{H}(t)|\Psi(t_{0})\right\rangle =\nonumber\\
 & \sum_{l}g_{kl}(t)\left\langle \Psi_{in}|\bar{\mathcal{A}}_{l}(t)|\Psi_{in}\right\rangle ,
 \end{align}
 \begin{align}
 & \left\langle \Psi(t_{0})|\mathcal{A}_{k}^{H}(t)\mathcal{A}_{l}^{H}(t')|\Psi(t_{0})\right\rangle =\nonumber\\
 &\sum_{l,j}g_{kl}(t)g_{ij}(t')\left\langle \Psi_{in})|\mathcal{\bar{A}}_{l}(t)\mathcal{\bar{A}}_{j}(t')|\Psi_{in}\right\rangle ,
\end{align}
and so on for higher order correlation functions, for arbitrary times
$t$ and $t'$. Since for $t>t_{max}$ we have $U(t,-\infty)=U(\infty,-\infty)$,
if $t$ and $t'$ are greater than $t_{max}$ we have 
\begin{align}
 & \left\langle \Psi(t_{0})|\mathcal{A}_{k}^{H}(t)|\Psi(t_{0})\right\rangle \rightarrow\sum_{l}g_{kl}(t)\left\langle \Psi_{in}|\bar{\mathcal{A}}_{l}(\infty)|\Psi_{in}\right\rangle ,
 \end{align}
 \begin{align}
 & \left\langle \Psi(t_{0})|\mathcal{A}_{k}^{H}(t)\mathcal{A}_{l}^{H}(t')|\Psi(t_{0})\right\rangle \rightarrow\nonumber\\
 &\sum_{l,j}g_{kl}(t)g_{ij}(t')\left\langle \Psi_{in})|\mathcal{\bar{A}}_{l}(\infty)\mathcal{\bar{A}}_{j}(\infty)|\Psi_{in}\right\rangle ,
\end{align}
and the remaining dynamics, governed only by $H_{0}$ is captured
by the $g_{ij}(t).$

\section{Dynamical equations}

In Appendix~\ref{appendix_dynamical} we show that 
\begin{align}
 & V\left(\left\{ \sum_{j}g_{ij}(t)\mathcal{\bar{A}}_{j}(t)\right\} \right)\label{eq:vresult}\\
 & =\hbar\sum_{J}\gamma_{J}^{*}\bar{b}_{J}(t)\bar{\psi}_{J}(-v_{J}t,t)+H.c.\nonumber \\
 & +\hbar\sum_{J}\gamma_{Jph}^{*}\bar{b}_{J}(t)\bar{\psi}_{Jph}(-v_{Jph}t,t)+H.c.\nonumber \\
 & +H_{NL}\left(\left\{ \bar{b}_{K}(t)e^{-i\omega_{K}t}\right\} ,\left\{ \bar{b}_{L}^{\dagger}(t)e^{i\omega_{L}t}\right\} \right),\nonumber 
\end{align}
where, for example, $\bar{\psi}(-v_{J}t,t)$, is the operator $\bar{\psi}(x,t)$
(see (\ref{eq:Abardef})) evaluated at $x=-v_{J}t$. Using this in
(\ref{eq:dynamical_equations}) we find the equation for $\bar{\psi}_{J}(x,t)$
is 
\begin{align}
 & i\hbar\frac{\partial\bar{\psi}_{J}(x,t)}{\partial t}=\hbar\gamma_{J}\bar{b}_{J}(t)\delta(x+v_{J}t),
\end{align}
and subject to the initial condition $\bar{\psi}_{J}(x,-\infty)=\psi_{J}(x)$
the solution is 
\begin{align}
 & \bar{\psi}_{J}(x,t)=\psi_{J}(x)-i\frac{\gamma_{J}}{v_{J}}\bar{b}_{J}\left(-\frac{x}{v_{J}}\right)\theta(t+\frac{x}{v_{J}}),
\end{align}
where as usual $\theta(t)=1,1/2,0$ as $t>0,=0,<0;$ thus
\begin{align}
 & \bar{\psi}_{J}(-v_{J}t,t)=\psi_{J}(-v_{J}t)-i\frac{\gamma_{J}}{2v_{J}}\bar{b}_{J}\left(t\right),
 \label{eq:psi_bar_out}
\end{align}
and similarly for $\bar{\psi}_{Jph}(-v_{Jph}t,t)$. Using these results
in the equation (\ref{eq:dynamical_equations}) for $\bar{b}_{J}(t)$
we find 
\begin{align}
 & \left(\frac{d}{dt}+\bar{\Gamma}_{J}\right)\bar{b}_{J}(t)=s_{J}(t) \nonumber \\
 &+\frac{1}{i\hbar}\left[\bar{b}_{J}(t),H_{NL}\left(\left\{ \bar{b}_{K}(t)e^{-i\omega_{K}t}\right\} ,\left\{ \bar{b}_{L}^{\dagger}(t)e^{i\omega_{L}t}\right\} \right)\right],\label{eq:work}
\end{align}
where the Schrödinger operator 
\begin{align}
 & s_{J}(t)=-i\gamma_{J}^{*}\psi_{J}(-v_{J}t)-i\gamma_{Jph}^{*}\psi_{Jph}(-v_{Jph}t),
\end{align}
and 
\begin{align}
 & \bar{\Gamma}_{J}\equiv\frac{\left|\gamma_{J}\right|^{2}}{2v_{J}}+\frac{\left|\gamma_{Jph}\right|^{2}}{2v_{Jph}}
\end{align}
describes the decay of the field in the ring due to coupling to
the channel and the scattering losses, the latter described by coupling to the phantom channel. 

\section{Operator dynamics}

For the resonances $L=p_{1},p_{2}$ associated with the two pumps we make
the classical approximation, taking $\psi_{L}(-v_{L}t)\rightarrow\left\langle \psi_{L}(-v_{L}t)\right\rangle $,
where in the center of long incident pump pulses we put
\begin{align}
 & \left\langle \psi_{L}(-v_{L}t)\right\rangle =\sqrt{\frac{P_{L}}{v_{L}\hbar\omega_{L}}}e^{-i\Delta_{L}t}\equiv C_{L}e^{-i\Delta_{L}t}
\end{align}
(cf. (\ref{eq:power})), defining a CW approximation; here $P_{L}$
is the incident power, and $\Delta_{L}$ a detuning from the ring
resonance at $\omega_{L}$. Similarly, for $L=p_{1},p_{2}$ we take
$\bar{b}_{L}(t)\rightarrow\left\langle \bar{b}_{L}(t)\right\rangle \equiv\beta_{L}(t).$
Then the classical limit of the equations (\ref{eq:work}) can be
constructed for $\beta_{p_{1}}(t)$ and $\beta_{p_{2}}(t)$, and introducing
\begin{equation}
F_{p_{1}}\left(t\right)\equiv\beta_{p_{1}}\left(t\right)e^{i\Delta_{p_{1}}t},\label{eq:Bp1_to_Fp1}
\end{equation}
\begin{equation}
F_{p_{2}}\left(t\right)\equiv\beta_{p_{2}}\left(t\right)e^{i\Delta_{p_{2}}t},\label{eq:Bm1_to_Fm1}
\end{equation}
they can be written as 
\begin{align}
& \left(\frac{d}{dt}+\bar{\Gamma}_{p_{1}}-i\Lambda\left(\left|F_{p_{1}}(t)\right|^{2}+2\left|F_{p_{2}}(t)\right|^{2}\right)-i\Delta_{p_{1}}\right)F_{p_{1}}(t) \nonumber \\
&=-i\gamma_{p_{1}}^{*}C_{p_{1}},\label{eq:F_p1}
\end{align}

\begin{align}
&\left(\frac{d}{dt}+\bar{\Gamma}_{p_{2}}-i\Lambda\left(\left|F_{p_{2}}(t)\right|^{2}+2\left|F_{p_{1}}(t)\right|^{2}\right)-i\Delta_{p_{2}}\right)F_{p_{2}}(t) \nonumber \\
&=-i\gamma_{p_{2}}^{*}C_{p_{2}},\label{eq:F_m1}
\end{align}
in which we have assumed there is no incoming pump energy in the phantom
channel. The steady state solutions of these equations are affected
by SPM and XPM, which determine the resonant frequencies of the
structure in the presence of the nonlinearity. We refer to these new
resonant frequencies as the ``hot-cavity resonances.'' Note that these two
equations can easily be solved numerically for the steady-state values of
$F_{p_{1}}$ and $F_{p_{2}}$, and one can use the transformations
(\ref{eq:Bp1_to_Fp1}) and (\ref{eq:Bm1_to_Fm1})
to determine the quantities $\beta_{p_{1}}\left(t\right)$ and $\beta_{p_{2}}\left(t\right)$.

With these in hand we can construct the equations for the fluctuating
quantities, 
\begin{align}
 & \tilde{b}_{J}(t)\equiv\bar{b}_{J}(t)-\left\langle \bar{b}_{J}(t)\right\rangle ,
\end{align}
which will be driven by the quantities 
\begin{align}
 & \tilde{s}_{J}(t)\equiv\bar{d}_{J}(t)-\left\langle \bar{d}_{J}(t)\right\rangle ,
\end{align}
where since $\left\langle \bar{s}_{J}(t)\right\rangle =0$ for $J=m,s,n$
we have $\left\langle \bar{b}_{J}(t)\right\rangle =0$ for those $J$.
For reasons we discuss below, it is convenient to work with the
quantities
\begin{equation}
f_{J}\left(t\right)\equiv\tilde{b}_{J}\left(t\right)e^{iR_{J}t},\label{eq:f_to_b}
\end{equation}
where 
\begin{align}
R_{s}=\frac{\Delta_{p_{1}}+\Delta_{p_{2}}}{2}+\frac{\omega_{p_{1}}+\omega_{p_{2}}-2\omega_{s}}{2},
\end{align}

\begin{align}
R_{p_{1}}=\Delta_{p_{1}},
\end{align}

\begin{align}
R_{p_{2}}=\Delta_{p_{2}},
\end{align}
\begin{align}
R_{m}=\frac{3\Delta_{p_{1}}-\Delta_{p_{2}}}{2}+\frac{3\omega_{p_{1}}-\omega_{p_{2}}-2\omega_{m}}{2},
\end{align}
\begin{align}
R_{n}=\frac{3\Delta_{p_{2}}-\Delta_{p_{1}}}{2}+\frac{3\omega_{p_{2}}-\omega_{p_{1}}-2\omega_{n}}{2}.
\end{align}
These are detuning-like quantities; in the limit where group-velocity dispersion is negligible and both pumps are equally detuned, $\Delta_{p_{2}}=\Delta_{p_{1}}=\Delta$, all the $R$ parameters are equal to that single detuning, $R_J = \Delta$.\par
The resulting dynamical equations, 
\begin{align}
 & \left(\frac{d}{dt}+\bar{\Gamma}_{s}-iR_{s}\right)f_{s}\left(t\right)=\tilde{s}_{s}\left(t\right)e^{iR_{s}t}\nonumber\\
 & +2i\Lambda\left(F_{p_{1}}\left(t\right)F_{p_{2}}\left(t\right)f_{s}^{\dagger}\left(t\right)+F_{p_{1}}^{*}\left(t\right)F_{p_{2}}\left(t\right)f_{m}\left(t\right)\right.\nonumber \\
 & \left.+F_{p_{2}}^{*}\left(t\right)F_{p_{1}}\left(t\right)f_{n}\left(t\right)+F_{p_{1}}^{*}\left(t\right)F_{p_{1}}\left(t\right)f_{s}\left(t\right)\right.\nonumber \\
 & \left. +F_{p_{2}}^{*}\left(t\right)F_{p_{2}}\left(t\right)f_{s}\left(t\right)\right)\nonumber \\
 & +i\Lambda\left(F_{p_{1}}\left(t\right)F_{p_{1}}\left(t\right)f_{m}^{\dagger}\left(t\right)+F_{p_{2}}\left(t\right)F_{p_{2}}\left(t\right)f_{n}^{\dagger}\left(t\right)\right),\label{eq:fs} 
\end{align}

\begin{align}
 & \left(\frac{d}{dt}+\bar{\Gamma}_{m}-iR_{m}\right)f_{m}\left(t\right)=\tilde{s}_{m}\left(t\right)e^{iR_{m}t}\nonumber \\
 & +2i\Lambda\left(F_{p_{2}}^{*}\left(t\right)F_{p_{1}}\left(t\right)f_{s}\left(t\right)+F_{p_{1}}^{*}\left(t\right)F_{p_{1}}\left(t\right)f_{m}\left(t\right)\right.\nonumber\\
 &\left.+F_{p_{2}}^{*}\left(t\right)F_{p_{2}}\left(t\right)f_{n}\left(t\right)\right)\nonumber\\
 & +i\Lambda\left(2F_{p_{1}}\left(t\right)F_{p_{2}}\left(t\right)f_{n}^{\dagger}\left(t\right)+F_{p_{1}}\left(t\right)F_{p_{1}}\left(t\right)f_{s}^{\dagger}\left(t\right)\right),\label{eq:fm} 
\end{align}

\begin{align}
 & \left(\frac{d}{dt}+\bar{\Gamma}_{n}-iR_{n}\right)f_{n}\left(t\right)=\tilde{s}_{n}\left(t\right)e^{iR_{n}t}\nonumber\\
 & +2i\Lambda\left(F_{p_{1}}^{*}\left(t\right)F_{p_{2}}\left(t\right)f_{s}\left(t\right)+F_{p_{1}}^{*}\left(t\right)F_{p_{1}}\left(t\right)f_{n}\left(t\right)\right.\nonumber\\
 &\left. +F_{p_{2}}^{*}\left(t\right)F_{p_{2}}\left(t\right)f_{n}\left(t\right)\right)\nonumber \\
 & +i\Lambda\left(2F_{p_{1}}\left(t\right)F_{p_{2}}\left(t\right)f_{m}^{\dagger}\left(t\right)+F_{p_{2}}\left(t\right)F_{p_{2}}\left(t\right)f_{s}^{\dagger}\left(t\right)\right),\label{eq:fn}
\end{align}
\begin{align}
 & \left(\frac{d}{dt}+\bar{\Gamma}_{p_{1}}-iR_{p_{1}}\right)f_{p_{1}}\left(t\right)=\tilde{s}_{p_{1}}\left(t\right)e^{iR_{p_{1}}t}\nonumber\\
 &+2i\Lambda\left(\left|F_{p_{2}}\left(t\right)\right|^{2}+\left|F_{p_{1}}\left(t\right)\right|^{2}\right)f_{p_{1}}\left(t\right)\nonumber\\
 & +2i\Lambda\,F_{p_{2}}^{*}\left(t\right)F_{p_{1}}\left(t\right)f_{p_{2}}\left(t\right)+2i\Lambda\,F_{p_{1}}\left(t\right)F_{p_{2}}\left(t\right)f_{p_{2}}^{\dagger}\left(t\right)\nonumber\\
 &+i\Lambda\,F_{p_{1}}^{2}\left(t\right)f_{p_{1}}^{\dagger}\left(t\right),\label{eq:fp1} 
\end{align}
\begin{align}
 & \left(\frac{d}{dt}+\bar{\Gamma}_{p_{2}}-iR_{p_{2}}\right)f_{p_{2}}\left(t\right)=\tilde{s}_{p_{2}}\left(t\right)e^{iR_{p_{2}}t}\nonumber\\
 &+2i\Lambda\left(\left|F_{p_{2}}\left(t\right)\right|^{2}+\left|F_{p_{1}}\left(t\right)\right|^{2}\right)f_{p_{2}}\left(t\right)\nonumber\\
 & +2i\Lambda\,F_{p_{1}}^{*}\left(t\right)F_{p_{2}}\left(t\right)f_{p_{1}}\left(t\right)+2i\Lambda\,F_{p_{2}}\left(t\right)F_{p_{1}}\left(t\right)f_{p_{1}}^{\dagger}\left(t\right)\nonumber\\
 &+i\Lambda\,F_{p_{2}}^{2}\left(t\right)f_{p_{2}}^{\dagger}\left(t\right),\label{eq:fp2} 
\end{align}
can be written in matrix form, 
\begin{equation}
\frac{d}{dt}\mathbb{\mathcal{F}}\left(t\right)=\mathbb{M}\left(t\right)\,\mathcal{F}\left(t\right)+\mathcal{D}\left(t\right),\label{eq:c_differential}
\end{equation}
where
\begin{equation}
\mathbb{\mathcal{F}}\left(t\right)=\left(\begin{array}{c}
\uparrow\\
f_{l}\left(t\right)\\
\downarrow\\
\uparrow\\
f_{l}^{\dagger}\left(t\right)\\
\downarrow
\end{array}\right),
\end{equation}
\begin{equation}
\mathcal{D}\left(t\right)=\left(\begin{array}{c}
\uparrow\\
\tilde{s}_{l}\left(t\right)e^{iR_{l}t}\\
\downarrow\\
\uparrow\\
\tilde{s}_{l}^{\dagger}\left(t\right)e^{-iR_{l}t}\\
\downarrow
\end{array}\right),
\end{equation}
and $\mathbb{M}\left(t\right)$ is a matrix, the components of which
can be read from (\ref{eq:fs},\ref{eq:fm},\ref{eq:fn},\ref{eq:fp1},\ref{eq:fp2}).
From the form (\ref{eq:c_differential}) we can construct a formal solution
by introducing a Green matrix $\mathbb{G}\left(t,t^{\prime}\right)$
satisfying
\begin{equation}
\frac{d\mathbb{G}\left(t,t^{\prime}\right)}{d\tau}=\mathbb{M}\left(t\right)\,\mathbb{G}\left(t,t^{\prime}\right),
\end{equation}
and $\mathbb{G}\left(t^{\prime},t^{\prime}\right)=\mathbb{I}$ for
all $t^{\prime}.$ We find the solution of (\ref{eq:c_differential}) as
\begin{equation}
\mathcal{F}\left(t\right)=\mathbb{G}\left(t,t_{0}\right)\,\mathcal{F}\left(t_{0}\right)+\int_{t_{0}}^{t}\mathbb{G}\left(t,t^{\prime}\right)\,\mathcal{D}\left(t^{\prime}\right)\,dt^{\prime}.
\end{equation}
We now consider $t_{0}\rightarrow-\infty$, and we can expect that
$\mathbb{G}\left(t,-\infty\right)$ vanishes because, due to coupling with the real and phantom channels, the initial state of the ring will be inconsequential at much later times. Therefore, we can write
\begin{equation}
\mathcal{F}\left(t\right)=\int_{-\infty}^{t}\mathbb{G}\left(t,t^{\prime}\right)\,\mathcal{D}\left(t^{\prime}\right)\,dt^{\prime}.
\end{equation}
The Green matrix can be written as
\begin{align}
\mathbb{G}\left(t,t^{\prime}\right)=\left(\begin{array}{cc}
G^{D}\left(t,t^{\prime}\right) & G^{C}\left(t,t^{\prime}\right)\\
G^{*C}\left(t,t^{\prime}\right) & G^{*D}\left(t,t^{\prime}\right)
\end{array}\right),
\end{align}
where each element itself is a $5\times5$ matrix. This allows us
to write

\begin{align}
f_{l}\left(t\right) & =\sum_{l^{\prime}}\int_{-\infty}^{t}G_{ll^{\prime}}^{D}\left(t,t^{\prime}\right)\,\tilde{s}_{l^{\prime}}\left(t^{\prime}\right)e^{iR_{l^{\prime}}t^{\prime}}\,dt^{\prime}\nonumber \\
 & +\sum_{l^{\prime}}\int_{-\infty}^{t}G_{ll^{\prime}}^{C}\left(t,t^{\prime}\right)\,\tilde{s}_{l^{\prime}}^{\dagger}\left(t^{\prime}\right)e^{-iR_{l^{\prime}}t^{\prime}}\,dt^{\prime},\label{eq:f_operator}
\end{align}
and correspondingly for $f_{l}^{\dagger}(t)$. Then using the commutation
relations 
\begin{equation}
\left[\tilde{s}_{l}\left(t^{\prime}\right),\tilde{s}_{l^{\prime}}\left(t^{\prime\prime}\right)\right]=0,\label{eq:commutation_d_d}
\end{equation}
\begin{equation}
\left[\tilde{s}_{l}\left(t^{\prime}\right),\tilde{s}_{l^{\prime}}^{\dagger}\left(t^{\prime\prime}\right)\right]=2\bar{\Gamma}_{l}\,\delta_{ll^{\prime}}\,\delta\left(t^{\prime}-t^{\prime\prime}\right),\label{eq:commutation_d_d_dagger}
\end{equation}
it is possible to evaluate the correlation functions $\left\langle \psi_{in}\left|f_{l}\left(t\right)f_{l^{\prime}}\left(t\right)\right|\psi_{in}\right\rangle $
and $\left\langle \psi_{in}\left|f_{l}^{\dagger}\left(t\right)f_{l^{\prime}}\left(t\right)\right|\psi_{in}\right\rangle ,$
and thus $\left\langle \psi_{in}\left|\tilde{b}_{l}\left(t\right)\tilde{b}_{l^{\prime}}\left(t\right)\right|\psi_{in}\right\rangle $
and $\left\langle \psi_{in}\left|\tilde{b}_{l}^{\dagger}\left(t\right)\tilde{b}_{l^{\prime}}\left(t\right)\right|\psi_{in}\right\rangle .$ We find
\begin{align}
 & \left\langle \psi_{in}\left|f_{J}\left(t_{2}\right)f_{J^{\prime}}\left(t_{1}\right)\right|\psi_{in}\right\rangle \nonumber \\
 & =2\,\Theta\left(t_{2}-t_{1}\right)\sum_{m}\bar{\Gamma}_{m}\,\int_{-\infty}^{t_{1}} G_{Jm}^{D}\left(t_{2},t^{\prime}\right)G_{J^{\prime}m}^{C}\left(t_{1},t^{\prime}\right)\,dt^{\prime}\nonumber \\
 & +2\,\Theta\left(t_{1}-t_{2}\right)\sum_{m}\bar{\Gamma}_{m}\,\int_{-\infty}^{t_{2}}G_{Jm}^{D}\left(t_{2},t^{\prime}\right)G_{J^{\prime}m}^{C}\left(t_{1},t^{\prime}\right)\,dt^{\prime},\label{eq:b1J_b1L}
\end{align}
and 
\begin{align}
 & \left\langle \psi_{in}\left|f_{J}^{\dagger}\left(t_{2}\right)f_{J^{\prime}}\left(t_{1}\right)\right|\psi_{in}\right\rangle \nonumber \\
 & =2\,\Theta\left(t_{2}-t_{1}\right)\sum_{m}\bar{\Gamma}_{m}\,\int_{-\infty}^{t_{1}}G_{Jm}^{*C}\left(t_{2},t^{\prime}\right)G_{J^{\prime}m}^{C}\left(t_{1},t^{\prime}\right)\,dt^{\prime}\nonumber \\
 & +2\,\Theta\left(t_{1}-t_{2}\right)\sum_{m}\bar{\Gamma}_{m}\,\int_{-\infty}^{t_{2}}G_{Jm}^{*C}\left(t_{2},t^{\prime}\right)G_{J^{\prime}m}^{C}\left(t_{1},t^{\prime}\right)\,dt^{\prime}.
 \label{eq:bd1J_b1L}
\end{align}
The reason for introducing the quantities $f_{l}(t)$ (\ref{eq:f_to_b})
is that, in the CW limit where $F_{p_1}(t)$ and $F_{p_2}(t)$ are independent
of time, the coefficients of the matrix $\mathbb{M}\left(t\right)$
are also independent of time. In that limit, when $\mathbb{M}\left(t\right) = \mathbb{M}$, the elements of the full Green matrix $\mathbb{G}\left(t,t^{\prime}\right)$
can be written as
\begin{equation}
\label{eq:G_in_eigens}
G_{ij}\left(t,t^{\prime}\right)=\sum_{k}V_{ik}\,e^{\lambda_{k}\left(t-t^{\prime}\right)}\,V_{kj}^{-1},
\end{equation}
where the $\lambda_{k}$ are the eigenvalues of $\mathbb{M}$, and the matrix element $V_{ik}$ is the $i^{th}$ component of the eigenvector associated with the $k^{th}$ eigenvalue.

\section{Results}
In a CW experiment the squeezing spectrum can be determined by mixing the output of channel $J$ with
a bright coherent local oscillator (LO) at frequency $\omega_{J}$, and with a tunable
phase, on a 50/50 beam splitter;
the outputs are then detected on two balanced fast photodiodes.
The difference photocurrent from these detectors is used to capture
the power spectral density of this photocurrent signal.
Writing the signal and the local oscillator electric fields as
\begin{align}
E\left(t\right)\propto\bar{\psi}_{J}\left(-v_{J}t,\infty\right)e^{-i\omega_{J}t}+\bar{\psi}_{J}^{\dagger}\left(-v_{J}t,\infty\right)e^{i\omega_{J}t}
\end{align}
and
\begin{align}
E_{LO}\left(t\right)\propto\alpha\,e^{-i\omega_{J}t+i\theta}+\alpha^*\,e^{i\omega_{J}t-i\theta},
\end{align}
 where the local oscillator is considered classical with amplitude $\alpha$,
for a fixed $\theta$ the squeezing spectrum is given by \cite{RevModPhys.82.1155, CRESSER198347},
\begin{align}
S\left(\Omega\right)=v_{g}\int_{-\infty}^{\infty}d\tau\,e^{-i\Omega\tau}\left\langle X_{\theta}(t)X_{\theta}(t+\tau)\right\rangle,
\end{align}
where 
\begin{align}
X_{\theta}\left(t\right)\equiv \bar{\psi}_{J}\left(-v_{J}t,\infty\right)e^{-i\theta}
+\bar{\psi}_{J}^{\dagger}\left(-v_{J}t,\infty\right)e^{i\theta}.\label{eq:X_het}
\end{align}
 In Appendix~\ref{appendix_squeezing_spectrum} we derive an analytic expression for $S\left(\Omega\right).$ 

For our sample calculations we consider the ring and waveguide, each with a cross-section of  $1500\, \text{nm} \times 800\, \text{nm}$ (see Fig.~\ref{fig:maximums_not_equal_sigma_appendix}), to be made of $\text{Si}_3\text{N}_4$, for which there is no two-photon absorption at telecommunication frequencies. We take the structure to be fully clad in $\text{SiO}_2$.  The ring is assumed to have radius $R=113\,\mu m$, with loaded and intrinsic quality factors of $Q=2\times 10^5$ and $Q_{int}=1\times 10^6$, respectively; these factors are achievable with current technology~\cite{Zhang2021, Arrazola2021}.   The nonlinear coefficient  $\Lambda=2\pi\times0.62\,\text{Hz}$ is calculated by simulating the mode profile distribution in the ring resonator using Lumerical's Mode Solutions.  We initially consider CW pump fields at $1550.8\,\text{nm}$ and $1553.6\,\text{nm}$, which in the ``cold cavity limit" -- i.e., when the microresonator is subject only to weak fields and nonlinear effects can be neglected -- correspond to pump fields at ring mode orders $n_J=830$ ($p_1$) and $n_J=832$ ($p_2$); we later envision tuning to get maximum output intensity by compensating the resonances' frequencies shift due to SPM and XPM. The vacuum wavelength of resonance $s$ in the cold cavity limit is $1551.9\,\text{nm}$, corresponding to ring mode order $n_J = 831$. 

Even in the cold cavity limit the frequency separations between adjacent resonances are not identical due to group-velocity dispersion. For instance, the frequency separation between the resonances $s$ and $p_2$ is $3\,\text{MHz}$ more than the separation between the resonances $s$ and $p_1$. However, this is so much smaller than the average of these two separations, which is  $0.2\,\text{THz}$, as well as so much smaller than the optically induced shifts and the resonance linewidths discussed below, that the results we present would not be significantly modified were group-velocity dispersion neglected in the calculation.

\begin{figure*}
\centering
\includegraphics[width=\textwidth]{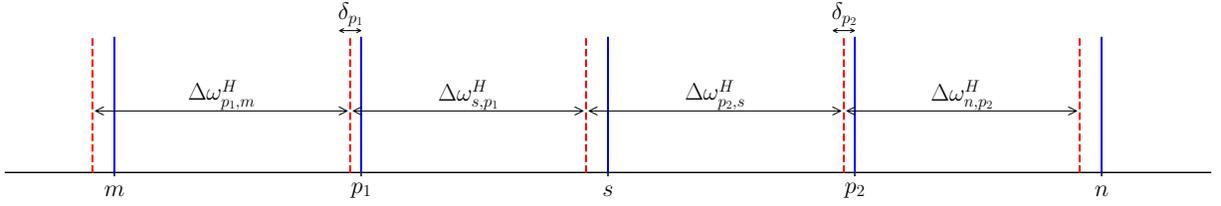}
\caption{(Color online) Illustration of the hot~(show in red) and cold~(shown in blue) cavity resonances. The resonance frequencies do not shift equally since the modes $p_1$ and $p_2$ are affected by SPM and XPM, while the modes $m$, $s$, and $n$ are affected only by XPM. Considering the the frequency shift caused by SPM to be $U$, the modes $p_1$ and $p_2$ move $3U$, and the other modes move $4U$ toward lower frequencies. The detuning of the pumps from the hot-cavity resonances are indicated by $\delta_{p_1}$ and $\delta_{p_2}$.}
\label{fig:hot_and_cold}
\end{figure*}

Optically induced shifts arise as the resonances $m$, $s$, and $n$ experience XPM from each of the pumps, while the resonances $p_1$ and $p_2$ are affected by both SPM and XPM.  By ``hot cavity" resonances we mean the cavity resonances in the presence of these nonlinear effects. We always consider equal pump powers in this paper, and so the frequency difference between the hot and cold cavity resonances will be $4U$ for the modes $m$, $s$, and $n$, and $3U$ for the modes $p_1$ and $p_2$, where $U=-\Lambda|F_{p_1}|^2=-\Lambda|F_{p_2}|^2$ is the frequency shift of each pump resonance due to SPM. As a result, the frequency separations of the resonances in the hot resonator, as shown schematically in Fig.~\ref{fig:hot_and_cold}, differ from each other. \par
 \par

In practice, the pump detunings and the input powers are the main parameters one can adjust to optimize the squeezing achieved in the output. Although increasing the input power results in enhancing DP-SFWM, the parasitic processes SP-SFWM and BS-FWM are enhanced as well. Increasing the input power can also lead to a transition into bistable or OPO regimes~\cite{Rukhlenko:10, PhysRevApplied.3.044005}. In this paper we avoid these regimes and only consider input powers for which the five-resonance approximation is valid, as confirmed by LLE simulations that show negligible intensity in the higher order sidebands at the pump intensities we consider.  \par
\begin{figure}
\includegraphics[width=\linewidth]{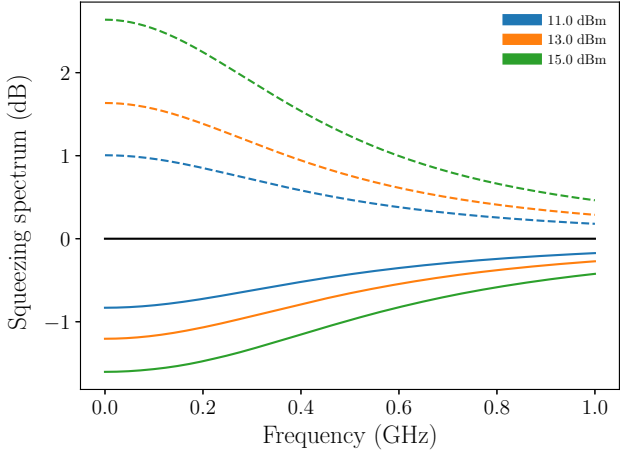}
\caption{(Color online) Squeezing~(solid lines) and anti-squeezing~(dashed lines) spectra for different input powers. The pumps are on resonance with the hot resonator. Black solid line is shot noise level.}
\label{fig:squeezing_input_power}
\end{figure}

In Fig.~\ref{fig:squeezing_input_power} we present the squeezing spectrum for different total input powers, where the pumps have equal powers; for each configuration the pumps are tuned to the hot cavity resonance. For the total input powers of 11, 13, 15 $\text{dBm}$, the hot-cavity detunings of the pumps are $-2\pi\times49.3$, $-2\pi\times76.3$, and $-2\pi\times122.5$ $\text{MHz}$, respectively, which are small fractions of the resonance linewidth  $2\bar{\Gamma}_s \approx 2\pi\times0.97\, \text{GHz}$. As can be seen, the highest degree of squeezing and anti-squeezing are achieved with the highest total input power. These calculations include the desired process, DP-SFWM, as well as all the others.

 \par
\begin{figure}
\includegraphics[width=\linewidth]{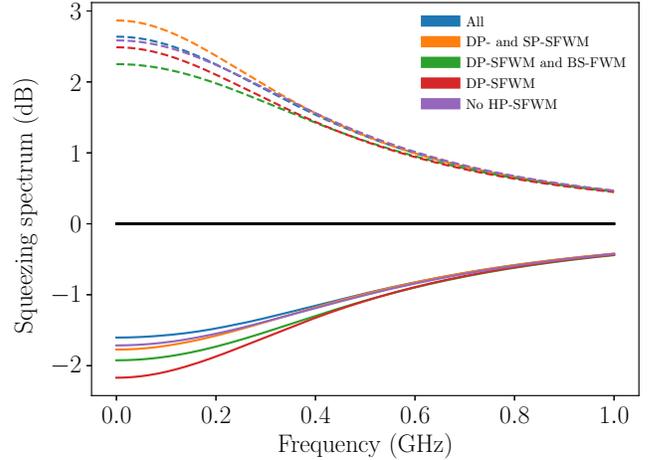}
\caption{(Color online) Squeezing~(solid lines) and anti-squeezing~(dashed lines) spectra for different combinations of nonlinear processes. Black solid line is shot noise level. The pumps are on resonance with the hot resonator. The highest squeezing is achieved when only DP-SFWM is included and the lowest is obtained when all the processes are involved. }
\label{fig:squeezing_processes}
\end{figure}

Our CMT analysis allows us to selectively turn on or off individual nonlinear processes, thus offering us an insight into their different roles.  In Fig.~\ref{fig:squeezing_processes} we plot the squeezing spectrum for different combinations of processes for a fixed input power, always keeping the two pumps at the hot cavity resonances and always including the DP-SFWM process. 

If only the DP-SFWM process is considered, we see that we achieve higher squeezing than any other circumstance, indicating that all the additional processes considered here are parasitic. The anti-squeezing is larger than the squeezing due to the scattering loss from the ring resonator.

If only the DP-SFWM and SP-SFWM processes are considered, we find that the squeezing decreases and the anti-squeezing increases from what would result if only DP-SFWM were present.  This arises because uncorrelated photons are injected into resonance $s$ by SP-SFWM (see Fig.~\ref{fig:processes}). On the other hand, if only the DP-SFWM and BS-FWM processes are considered, we find that both the squeezing and the anti-squeezing decrease from what would result if only DP-SFWM were present. This arises because photons are removed from resonance $s$ by BS-FWM (see Fig.~\ref{fig:processes}), and although the associated noise leads to a decrease in the squeezing, the removal of photons leads to a decrease in the anti-squeezing as well.  Finally, we see that if we include all processes except HP-SFWM, and then compare with the situation when all processes are included, we find a small decrease in squeezing and a small increase in anti-squeezing when HP-SFWM is included.  This arises because HP-SFWM injects photons in resonances $m$ and $n$ (see Fig.~\ref{fig:processes}), which then can lead to photons being injected into resonance $s$ through BS-FWM, analogous to the effect of SP-SFWM.

 \par
%Before discussing the effect of SPM and XPM on the squeezing spectrum, it is worth differentiating between the classical and quantum self and cross phase modulations. The former is meant when the two pumps are considered classically. Since the classical pumps have high enough powers, they are responsible for any detuning from the cold cavity and their effect cannot be neglected. The highest intensity in the ring is achieved by overcoming the resulting detuning from SPM and XPM via tuning the pumps to be on resonance with the hot cavity. The latter, on the other hand, is weak and does not change the intensity inside the ring significantly. However, it is clear that due to the group velocity dispersion in the ring, the frequency separation between the ring resonances are not equal. This implies that the average frequency of the two pumps do not exactly locate on the mode $s$ and hence there will be a detuning there. When considering the quantum SPM and XPM, the resonance frequency of the mode $s$ shifts towards the average frequency of the two pumps and enhances the squeezing comparing to the case where DP-SFWM is only included.\par
%
Of the parasitic processes, SP-SFWM is the most effective in reducing the squeezing, justifying efforts to design a structure to suppress this unwanted process without significantly compromising the generation efficiency~\cite{Zhang2021}. Even without modifying the structure, e.g., by adding an auxiliary ring~\cite{Zhang2021}, and without relying on materials or designs yielding larger group velocity dispersion so that dispersion engineering can be employed~\cite{Kippenbergeaan8083,Kim2017}, pump detuning can be be employed as a parameter to affect the squeezing achieved in the output, and when chosen properly can lead to enhanced squeezing by suppressing the unwanted processes. To enhance the squeezing generated in the resonance $s$, the effectiveness of the SP-SFWM process should be degraded while keeping the DP-SFWM process as effective as possible. 

\begin{figure}
\includegraphics[width=\linewidth]{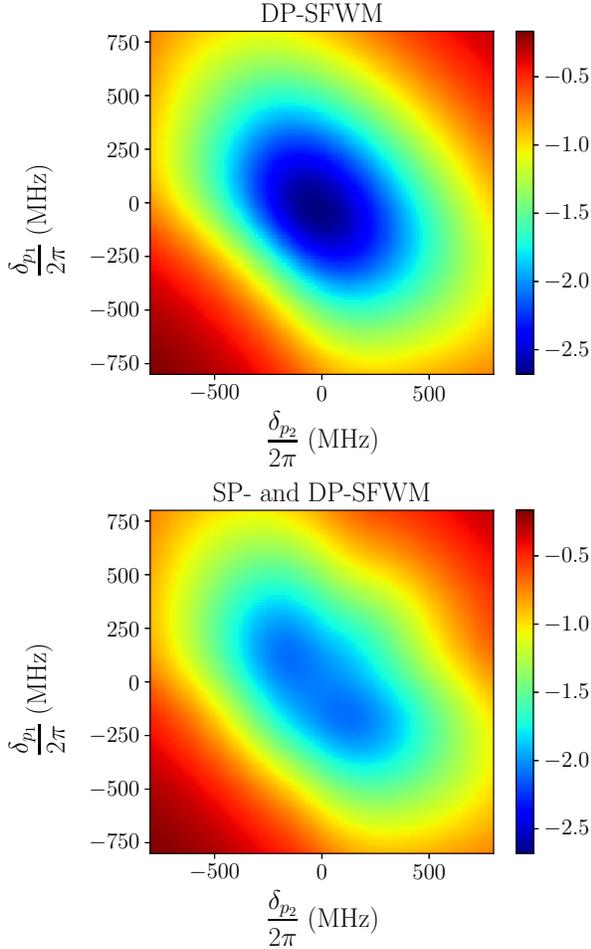}
\caption{(Color online) Squeezing in dB (top) when only DP-SFWM is included and (bottom) when both SP- and DP-SFWM are included as functions of the detunings of the first and the second pumps from the hot cavity resonances. Both cases have been simulated using the same total input power of $16~\text{dBm}$.  }
\label{fig:2D_16dBm}
\end{figure}

This can be achieved by breaking the frequency matching condition for SP-SFWM, while keeping that for DP-SFWM. As can be seen in Figs.~\ref{fig:processes} and \ref{fig:hot_and_cold}, this can be done by introducing $\delta_{p_1}$ and $\delta_{p_2}$ as detunings of the first and second pumps from the hot cavity resonances, respectively. To illustrate the effects of these pump detunings on the squeezing achieved in resonance $s$, in Fig.~\ref{fig:2D_16dBm} we plot the squeezing in that resonance as a function of $\delta_{p_1}$ and $\delta_{p_2}$, for a fixed total input power of 16 dBm. Focusing on SP-SFWM as the main parasitic process degrading the squeezing, we plot the generated squeezing in resonance $s$ considering only DP-SFWM, and considering both DP-SFWM and SP-SFWM.  When only DP-SFWM is considered, the highest squeezing is achieved where the two pumps tuned very close to the hot cavity resonances, $\delta_{p_1}=\delta_{p_2}=0$; the maximum squeezing actually occurs when the detunings are negative and on the order of $2\pi \times10\,\text{MHz}$, due to both SPM/XPM effects (see Fig.~\ref{fig:hot_and_cold}) and group velocity dispersion. However, when SP-SFWM processes are included the highest amount of squeezing occurs far from $\delta_{p_1}=\delta_{p_2}=0$. There are two local minima in the squeezing  at points very close to the symmetric detuning line, $\delta_{p_1}=-\delta_{p_2}$. Here energy conservation still allows DP-SFWM, but forbids SP-SFWM. Yet while such a detuning configuration suppresses the SP-SFWM processes, it also changes the energy of the pump fields in the ring, $E_{p_1}$ and $E_{p_2}$ respectively for the first and second pump, and hence decreases the number of generated correlated photons in resonance $s$ via DP-SPWM. Note that the close proximity of the two minima to the symmetric detuning line arises because of the low group velocity dispersion of the structure; this would hold generally over a few resonances for $\text{Si}_3 \text{N}_4$ structures, but would be violated for wide frequency spans, and indeed for narrow frequency spans in structures made of other materials, such as Si.\par
\begin{figure}
\includegraphics[width=\linewidth]{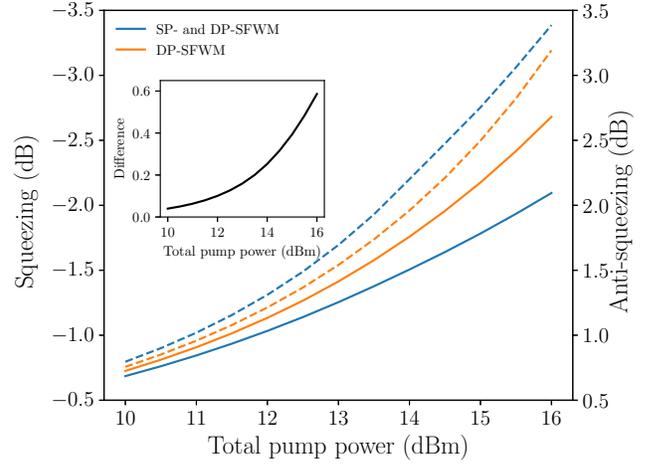}
\caption{(Color online) Highest squeezing~(solid lines) achievable for different pump powers with and without SP-SFWM. The values have been obtained by searching for the highest values in the 2D detuning space (similar to Fig.~\ref{fig:2D_16dBm}) for each total input power in the range; we also plot the value of the anti-squeezing (dotted lines) under these conditions. The inset shows the difference between the highest squeezing when including SP-SFWM and the case when only DP-SFWM is considered. }
\label{fig:different_power_sq_Asq}
\end{figure}

To emphasize the role of SP-SFWM on squeezing, in Fig.~\ref{fig:different_power_sq_Asq} we plot the highest squeezing achievable by adjusting the pump detunings for a range of input powers, with and without including SP-SFWM; we also plot the anti-squeezing under these conditions. Clearly in both instances the squeezing improves as the total input power is increased. However, the parasitic effect of SP-SFWM becomes more significant at higher pump powers; as shown in the inset of Fig.~\ref{fig:different_power_sq_Asq}, the difference in the maximum squeezing with and without SP-SFWM becomes more significant as the power of the pumps is increased. The result is that a higher pump power penalty is required to compensate for the parasitic effects of SP-SFWM as the level of desired squeezing is raised. \par    
\begin{figure}[!htb]
\includegraphics[width=\linewidth]{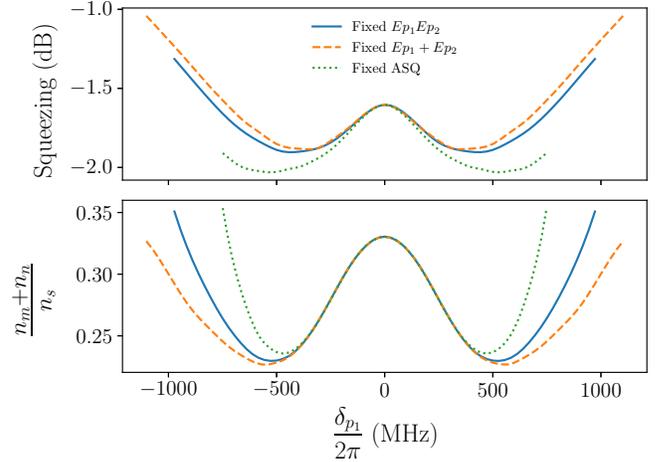}
\caption{(Color online) Squeezing~(top) and ratio of the total generated photons in the modes $m$ and $n$ to the number of generated photons in the mode $s$~(bottom) as a function of the detuning of the first pump from the hot-cavity resonance for three different scenarios of fixing the product of the energies at $69.6\,\text{pJ}^2$~(solid lines), total energy in the ring at $16.7\,\text{pJ}$~(dashed lines), and anti-squeezing at $2.64\,\text{dB}$~(dotted lines). The pumps are detuned symmetrically, i.e., $\delta_{p_1} = -\delta_{p_2}$, and the calculations included DP-SFWM and SP-SFWM. }
\label{fig:squeezing_detuning}
\end{figure}

 If one accepts this penalty, then significant improvements in squeezing can be achieved by detuning. Restricting our calculations to the main processes, DP-SFWM and SP-SFWM, in Fig.~\ref{fig:squeezing_detuning} we plot the level of squeezing that can be achieved keeping the product $E_{p_1}E_{p_2}$ fixed by adjusting the input powers, as different symmetric detunings from the hot cavity resonance -- which to good approximation yield the best values of squeezing -- are considered. The results presented in Fig.~\ref{fig:squeezing_detuning} with solid lines are obtained by setting $E_{p_1}E_{p_2}=69.6\,\text{pJ}^2$, which is the product of the energies for the total input power at $15\,\text{dBm}$ when the pumps are both tuned to the hot-cavity resonances. As the product of powers is kept fixed and symmetric detuning is introduced, the total input power ranges up to $21.5\,\text{dBm}$ over the range of detuning shown in Fig.~\ref{fig:2D_16dBm}. Moving away from the center ($\delta_{p_1} = 0$), there is a significant enhancement in squeezing, which is the result of suppressing the SP-SFWM while keeping the effect of DP-SFWM essentially unchanged. To confirm this, in  Fig.~\ref{fig:squeezing_detuning} we also plot the corresponding ratio of the total number of generated photons in the modes $m$ and $n$ to the number of photons in the mode $s$. Calculating the squeezing while holding fixed the total energy in the ring, $E_{p_1}+E_{p_2}$, or the anti-squeezing in the mode $s$, leads to similar improvements in the squeezing, as shown in the top diagram of   Fig.~\ref{fig:squeezing_detuning}, linked to the same behavior of the photon-number ratio, as shown in the bottom of that figure.             
 \par

\section{conclusion}
We have studied the generation of squeezed vacuum states via spontaneous four-wave mixing (SFWM) in a dual-pumped nonlinear ring resonator, taking into account scattering losses in the ring and the plethora of third order nonlinear optical processes. We applied coupled-mode theory and only considered five resonances, which is a reasonable approximation for the range of input pump powers considered. This approach allowed us to investigate the role of different processes by including or omitting them in the calculation, which is not possible in calculations based on the Lugiato-Lefever equations (LLE). Including self- and cross-phase modulation allowed us to take into account hot-cavity frequency shifts. We studied CW excitation and employed a Green function strategy to calculate the relevant physical quantities, such as squeezing and the number of generated photons. We showed that parasitic processes generate uncorrelated photons in the resonance of interest, and therefore degrade the quality of squeezing in the output; the most important of these parasitic processes is single-pump SFWM.  We demonstrated that without changing the ring resonator structure the use of symmetric pump detuning can suppress these unwanted processes, in particular by breaking the energy conservation condition for single-pump SFWM but not that for dual-pump SFWM. However, a power penalty must be accepted, because the dual-pump SFWM is then detuned from resonance. We intend to present LLE calculations, which are valid at higher pump powers than the five-resonance calculation used here, and studies in the regime of pulsed excitation, in future work.  

Finally, it is worth mentioning that the parameters considered in this work, while achievable and indeed typical, have not been optimised to provide maximum squeezing generation efficiency. Moreover, it is possible to combine the pump-detuning technique with dispersion engineering to suppress the parasitic processes more efficiently in a dual-pumped single ring resonator configuration. 

\section*{Acknowledgement}
This work is supported by Mitacs through the Mitacs Elevate program grant IT15614. J.E.S. acknowledges support from the Natural Sciences and Engineering Research Council of Canada (NSERC).
\appendix
\section{}
\label{Lambda_appendix}

Here we reduce the expression (\ref{Lambda_JKLM_1}) for $\Lambda_{JKLM}$ in the limit
where all frequencies are very close and we can take $\Lambda_{JKLM}\rightarrow\Lambda$.
In this limit we use $\omega$ to denote a typical ring frequency
($\omega_{J},\omega_{K},\omega_{L},$ or $\omega_{M}$), $\mathsf{d}(\boldsymbol{r}_{\perp};\zeta$)
to denote a typical mode amplitude, and $v$ to denote the group velocity
of light in the ring at frequency $\omega$. The expression (\ref{eq:gamma_def}) for
$\Gamma_{(3)}^{ijkl}(\boldsymbol{r})$ neglects material dispersion,
but we take the $n(\boldsymbol{r})$ there to be $n(\boldsymbol{r}_{\perp};\omega)$,
and use the relation
\begin{align}
 & \mathsf{d}(\boldsymbol{r}_{\perp};\zeta)=\epsilon_{0}n^{2}(\boldsymbol{r}_{\perp};\omega)\mathsf{e}(\boldsymbol{r}_{\perp};\zeta),\label{eq:dtoe}
\end{align}
where $\mathsf{e}(\boldsymbol{r}_{\perp};\zeta)$ is the electric
field mode amplitude, to write (\ref{Lambda_JKLM_1}) as 
\begin{align*}
 & \Lambda=\frac{3\hbar\omega^{2}\epsilon_{0}}{4\mathcal{L}^{2}}\times\\
 &\int d\boldsymbol{r}_{\perp}d\zeta\:\chi_{(3)}^{ijkl}(\boldsymbol{r}_{\perp})\left(\mathsf{e}^{i}(\boldsymbol{r}_{\perp};\zeta)\mathsf{e}^{j}(\boldsymbol{r}_{\perp};\zeta)\right)^{*}\mathsf{e}^{k}(\boldsymbol{r}_{\perp};\zeta)\mathsf{e}^{l}(\boldsymbol{r}_{\perp};\zeta),
\end{align*}
or 
\begin{align*}
 & \Lambda=\frac{\hbar\omega v^{2}\gamma}{\mathcal{L}},
\end{align*}
where the parameter $\gamma$ is given by 
\begin{align*}
 & \gamma=\frac{3\omega^{2}\epsilon_{0}}{4v^{2}\mathcal{L}}\times\\
 &\int d\boldsymbol{r}_{\perp}d\zeta\:\chi_{(3)}^{ijkl}(\boldsymbol{r}_{\perp})\left(\mathsf{e}^{i}(\boldsymbol{r}_{\perp};\zeta)\mathsf{e}^{j}(\boldsymbol{r}_{\perp};\zeta)\right)^{*}\mathsf{e}^{k}(\boldsymbol{r}_{\perp};\zeta)\mathsf{e}^{l}(\boldsymbol{r}_{\perp};\zeta).
\end{align*}
Introducing typical values $\overline{\chi}_{(3)}$ and $\overline{n}$
respectively for the elements of $\chi_{(3)}^{ijkl}$ and $n(\boldsymbol{r}_{\perp};\omega)$
in the region where the field amplitudes are concentrated, we can
write 
\begin{align}
 & \gamma=\frac{3\omega\overline{\chi}_{(3)}}{\overline{n}^{2}\epsilon_{0}c^{2}}\frac{1}{A},\label{eq:gamma}
\end{align}
where 
\begin{widetext}
\begin{align}
 & \frac{1}{A}=\frac{\mathcal{L}^{-1}\int d\boldsymbol{r}_{\perp}d\zeta\:\left(\chi_{(3)}^{ijkl}(\boldsymbol{r}_{\perp})/\overline{\chi}_{(3)}\right)\left(\mathsf{e}^{i}(\boldsymbol{r}_{\perp};\zeta)\mathsf{e}^{j}(\boldsymbol{r}_{\perp};\zeta)\right)^{*}\mathsf{e}^{k}(\boldsymbol{r}_{\perp};\zeta)\mathsf{e}^{l}(\boldsymbol{r}_{\perp};\zeta)}{\left(\int\mathsf{e}^{*}(\boldsymbol{r}_{\perp};0)\cdot\mathsf{e}(\boldsymbol{r}_{\perp};0)\frac{n(\boldsymbol{r}_{\perp};\omega)/\bar{n}}{v_{g}(\boldsymbol{r}_{\perp};\omega)/v}d\boldsymbol{r}_{\perp}\right)^{2}},\label{eq:OOA}
\end{align}
\end{widetext}
and we have used the normalization condition (\ref{eq:normalization}) together with (\ref{eq:dtoe});
thus (\ref{eq:OOA}) can be evaluated using electric field amplitudes
that are not normalized. Here $A$ has units of area, which we see
below can be identified as the effective area of the ring mode~\cite{boyd2019nonlinear}.\par
The general form of the electric field mode amplitude $\mathsf{e}(\boldsymbol{r}_{\perp};\zeta)$ can be written as

\begin{align}
 \mathsf{e}(\boldsymbol{r}_{\perp};\zeta)=&\left(\hat{\boldsymbol{x}}\sin\phi-\hat{\boldsymbol{y}}\cos\phi\right)e_{\rho}(\boldsymbol{r}_{\perp})\nonumber\\
 &+\left(\hat{\boldsymbol{x}}\cos\phi+\hat{\boldsymbol{y}}\sin\phi\right)e_{\phi}(\boldsymbol{r}_{\perp})\nonumber\\
 &+\hat{\boldsymbol{z}}e_{z}(\boldsymbol{r}_{\perp}),\label{eq:general_mode_at_phi}
\end{align}
where recall $\phi=\zeta/R$. It is worth considering two special cases, where the electric field is mainly in the plane of the wafer (normal to the wafer), known as TE (TM) ring modes. For a TM ring mode, to a good approximation, we can take 
\begin{align}
 & \mathsf{e}(\boldsymbol{r}_{\perp};\zeta)=\hat{\boldsymbol{z}}e_{z}(\boldsymbol{r}_{\perp}).\label{eq:TM}
\end{align}
Then, choosing $\overline{\chi}_{(3)}=\chi_{(3)}^{zzzz},$ and using
the approximation 
\begin{align}
 & \frac{n(\boldsymbol{r}_{\perp};\omega)/\bar{n}}{v_{g}(\boldsymbol{r}_{\perp};\omega)/v}\approx1\label{eq:approx}
\end{align}
in the region where the fields are concentrated, we find 
\begin{align}
 & \frac{1}{A}=\frac{\int d\boldsymbol{r}_{\perp}\left|e(\boldsymbol{r}_{\perp})\right|^{4}}{\left(\int d\boldsymbol{r}_{\perp}\left|e(\boldsymbol{r}_{\perp})\right|^{2}\right)^{2}},\label{eq:OOAresult}
\end{align}
where $e(\boldsymbol{r}_{\perp}) = e_{z}(\boldsymbol{r}_{\perp})$.
If instead we consider a TE ring mode, for which to good approximation
we can take 
\begin{align}
 & \mathsf{e}(\boldsymbol{r}_{\perp};\zeta)=\left(\hat{\boldsymbol{x}}\sin\phi-\hat{\boldsymbol{y}}\cos\phi\right)e_{\rho}(\boldsymbol{r}_{\perp}).\label{eq:TE}
\end{align}
Then assuming an isotropic material,
for which $\chi_{(3)}^{xxxx}=\chi_{(3)}^{yyyy}=\chi_{(3)}^{zzzz}$
and $\chi_{(3)}^{xxyy}=\chi_{(3)}^{zzzz}/3$ (and correspondingly
for all other tensor components with two $x$ indices and two $y$
indices), and using again the approximation (\ref{eq:approx}), we
also find (\ref{eq:OOAresult}), with now $e(\boldsymbol{r}_{\perp}) = e_{\rho}(\boldsymbol{r}_{\perp})$, as might be expected. Of course,
the $e_{\rho}(\boldsymbol{r}_{\perp})$ of (\ref{eq:TE}) need not be the
same as the $e_z(\boldsymbol{r}_{\perp})$ of (\ref{eq:TM}).

Using (\ref{eq:OOAresult}) in (\ref{eq:gamma}) we see that
since $A$ is the effective area of the mode, within the usual approximations
we have made here $\gamma$ is the standard result for the waveguide
nonlinear parameter of an optical fiber or channel waveguide~\cite{boyd2019nonlinear}. Naturally, the expression (\ref{eq:OOA}) for $A$ can
be evaluated numerically using (\ref{eq:general_mode_at_phi}) instead of (\ref{eq:TE}) or (\ref{eq:TM}), and without the use of the approximations (\ref{eq:approx}), and the value of $\gamma$
and thus of $\Lambda$ determined directly. This is in fact what we do in Sec. IX.

\section{}
\label{appendix_dynamical}

Here we detail some of the steps in the derivation of (\ref{eq:vresult}).
We first note that taking the time derivative of 
\begin{align}
 & \psi_{J}^{o}(x,t)\equiv\mathcal{U}_{0}^{\dagger}(t,0)\psi_{J}(x)\mathcal{U}_{0}(t,0),
\end{align}
 using (\ref{eq:Hchanneluse}), leads to the equation 
\begin{align}
 & \frac{\partial\psi_{J}^{o}(x,t)}{\partial t}+v_{J}\frac{\partial\psi_{J}^{o}(x,t)}{\partial x}=-i\omega_{J}\psi_{J}^{o}(x,t),
\end{align}
the solutions of which are 
\begin{align}
 & \psi_{J}^{o}(x,t)=e^{-i\omega_{J}t}\psi_{J}^{o}(x-v_{J}t,0)=e^{-i\omega_{J}t}\psi_{J}(x-v_{J}t).\label{eq:channel_prop}
\end{align}
Thus we can write 
\begin{align}
 & \mathcal{U}_{0}^{\dagger}(t,0)\psi_{J}(x)\mathcal{U}_{0}(t,0)=e^{-i\omega_{J}t}\psi_{J}(x-v_{J}t)\label{eq:gwork}\\
 & =\int g(x,x';t)\psi_{J}(x')dx',\nonumber 
\end{align}
where we have defined 
\begin{align}
 & g(x,x';t)=e^{-i\omega_{J}t}\delta(x'-x+v_{J}t),
\end{align}
so (\ref{eq:gwork}) is the immediate generalization of (\ref{eq:gdef}).
A corresponding result holds for the phantom channel field, and since
we immediately have 
\begin{align}
 & \mathcal{U}_{0}^{\dagger}(t,0)b_{J}\mathcal{U}_{0}(t,0)=e^{-i\omega_{J}t}b_{J},
\end{align}
using the definition (\ref{eq:Hamiltonian_start}) of $V(\left\{ \mathcal{A}_{i}\right\} t)$
we have, for example,
\begin{align}
 & \mathcal{U}_{0}^{\dagger}(t,0)V\left(\left\{ \mathcal{A}_{i}\right\} ;t\right)\mathcal{U}_{0}(t,0)=V\left(\left\{ \sum_{j}g_{ij}(t)\mathcal{A}_{j}\right\} \right)\label{eq:vresultbaby}\\
 & =\hbar\sum_{J}\gamma_{J}^{*}b_{J}\psi_{J}(-v_{J}t)+H.c.\nonumber \\
 & +\hbar\sum_{J}\gamma_{Jph}^{*}b_{J}\psi_{Jph}(-v_{Jph}t)+H.c.\nonumber \\
 & +H_{NL}\left(\left\{ b_{J}e^{-i\omega_{J}t}\right\} ,\left\{ b_{L}^{\dagger}e^{i\omega_{L}t}\right\} \right),\nonumber 
\end{align}
where $\psi_{J}(-v_{J}t)$, for example, is the Schrödinger operator
$\psi_{J}(x)$ evaluated at $x=-v_{J}t$. We require not this but
$V\left(\left\{ \sum_{j}g_{ij}(t)\mathcal{\bar{A}}_{j}(t)\right\} \right)$;
nonetheless, noting that (\ref{eq:vresultbaby}) can be taken to simply
identify the $g_{ij}(t),$ (\ref{eq:vresult}) then follows.

\section{}
\label{appendix_squeezing_spectrum}
Here we derive an analytic expression for $S\left(\Omega\right)$ in a case where $| \psi_{in}\rangle$ is the vacuum state; this is the situation of interest here. Using Eq.~(\ref{eq:psi_bar_out}) in this special case we find 
\begin{align}
 & \left\langle \bar{\psi}_{J}^{\dagger}(-v_{J}t_{2},\infty)\bar{\psi}_{J}(-v_{J}t_{1},\infty)\right\rangle =\frac{\left|\gamma_{J}\right|^{2}}{v_{J}^{2}}\left\langle \tilde{b}_{J}^{\dagger}(t_{2})\tilde{b}_{J}(t_{1})\right\rangle ,\label{eq:psi_dagger_psi_vacuum}
\end{align}
and 
\begin{align}
 & \left\langle \bar{\psi}_{J}(-v_{J}t_{2},\infty)\bar{\psi}_{J}(-v_{J}t_{1},\infty)\right\rangle = \nonumber \\
 & -\frac{\gamma_{J}^{2}}{v_{J}^{2}}\left(\left\langle \tilde{b}_{J}(t_{2})\tilde{b}_{J}(t_{1})\right\rangle \Theta(t_{2}-t_{1})\right.\nonumber\\
 & \left.  +\left\langle \tilde{b}_{J}(t_{1})\tilde{b}_{J}(t_{2})\right\rangle \Theta(t_{1}-t_{2})\right).
\end{align}
By employing Eq.~(\ref{eq:G_in_eigens}) in Eqs.~(\ref{eq:b1J_b1L}) and (\ref{eq:bd1J_b1L}) we find the required expectation values in terms of the eigenvalues and eigenvectors of $\mathbb{M}\left(t\right)$ as
\begin{align}
 & \left\langle \psi_{in}\left|\tilde{b}_{J}\left(t_{2}\right)\tilde{b}_{J}\left(t_{1}\right)\right|\psi_{in}\right\rangle \nonumber \\
 & =-2\,\sum_{m=1}^{N}\sum_{k^{\prime}=1}^{2N}\sum_{k^{\prime\prime}=1}^{2N}\bar{\Gamma}_{m}\frac{V_{J\,k^{\prime}}\,V_{k^{\prime}\,m}^{-1}\,V_{L\,k^{\prime\prime}}\,V_{k^{\prime\prime}\,m+N}^{-1}}{\lambda_{k^{\prime}}+\lambda_{k^{\prime\prime}}} \times \nonumber \\
 & \left(e^{\lambda_{k^{\prime}}\left(t_{2}-t_{1}\right)}\Theta\left(t_{2}-t_{1}\right)+e^{\lambda_{k^{\prime\prime}}\left(t_{1}-t_{2}\right)}\Theta\left(t_{1}-t_{2}\right)\right)\
\end{align}
and
\begin{align}
 & \left\langle \psi_{in}\left|\tilde{b}_{J}^{\dagger}\left(t_{2}\right)\tilde{b}_{J}\left(t_{1}\right)\right|\psi_{in}\right\rangle \nonumber\\
 & =-2\,\sum_{m=1}^{N}\sum_{k^{\prime}=1}^{2N}\sum_{k^{\prime\prime}=1}^{2N}\bar{\Gamma}_{m}\frac{V_{J+N\,k^{\prime}}\,V_{k^{\prime}\,m}^{-1}\,V_{J\,k^{\prime\prime}}\,V_{k^{\prime\prime}\,m+N}^{-1}}{\lambda_{k^{\prime}}+\lambda_{k^{\prime\prime}}}\times \nonumber\\
 & \left(e^{\lambda_{k^{\prime}}\left(t_{2}-t_{1}\right)}\Theta\left(t_{2}-t_{1}\right)+e^{\lambda_{k^{\prime\prime}}\left(t_{1}-t_{2}\right)}\Theta\left(t_{1}-t_{2}\right)\right),
\end{align}
where $N=5$ when only considering five modes. Letting $t_1 = t_2 + \tau$, and using the equations above, we can finally derive
\begin{align}
 & S_{J}\left(\Omega\right)=1-\sum_{m=1}^{N}\sum_{k^{\prime}=1}^{2N}\sum_{k^{\prime\prime}=1}^{2N}\frac{\bar{\Gamma}_{m}}{v_{g}}V_{k^{\prime}\,m}^{-1}\,V_{k^{\prime\prime}\,m+N}^{-1}\times\nonumber\\
 & \Big(e^{-2i\theta}\,\gamma_{J}^{2}\,V_{J\,k^{\prime}}\,V_{J\,k^{\prime\prime}}\,A_{k^{\prime},k^{\prime\prime}}\nonumber\\
 & -\left|\gamma_{J}\right|^{2}\,V_{J+N\,k^{\prime}}\,V_{J\,k^{\prime\prime}}\,\left(A_{k^{\prime},k^{\prime\prime}}+A_{k^{\prime\prime},k^{\prime}}\right)\nonumber\\
 & +e^{2i\theta}\,\left(\gamma_{J}^{*}\right)^{2}V_{J+N\,k^{\prime}}\,V_{J+N\,k^{\prime\prime}}\,A_{k^{\prime\prime},k^{\prime}}\Big),
\end{align}
where 
\begin{align}
    A_{i,j} \equiv \frac{4\lambda_{i}}{\left(\lambda_{i}+\lambda_{j}\right)\left(\lambda_{i}^{2}+\Omega^{2}\right)}
\end{align}
\bibliography{apssamp}% Produces the bibliography via BibTeX.

\end{document}